\documentclass[3p]{elsarticle}

\usepackage{mydefs}
\begin{document}

\begin{frontmatter}
%
% put the title of your paper here
%
% \title{Designing Qusecs Optimally: Quasi-uniform or Self-similar Layered Materials as Recursively Decomposed Solutions of Geometric Constraint Systems}

\title{Optimal Decomposition and Recombination of Isostatic Geometric Constraint Systems for Designing Layered Materials}

%
% put your submission number here
% this number will appear instead of authors and affiliations if the class option "submission" is active
%
% \SubNumber{110}

%
% put the author names here
% use the second argument as a reference to the list of affiliations
% no authors and affiliations will appear if the class option "submission" is active
%
% \author{Troy Baker}{1}
% \author{Meera Sitharam}{1}
% \author{Menghan Wang}{1}
% \author{Joel Willoughby}{1}

% \author{Troy Baker}
% \author{Meera Sitharam}
% \author{Menghan Wang}
% \author{Joel Willoughby}

\author[uf]{T.~Baker\corref{cor}}
\ead{tbaker@cise.ufl.edu}

\author[uf]{M.~Sitharam}
\ead{sitharam@cise.ufl.edu}

\author[uf]{M.~Wang}
\ead{menghan@cise.ufl.edu}

\author[uf]{J.~Willoughby}
\ead{jw5@cise.ufl.edu}

\address[uf]{University of Florida, CSE Building, Gainesville, Florida, U.S.A., 32611}

\cortext[cor]{Corresponding author}

%
% put the affiliations of the authors here
%
% \affiliation{1}{Department of Computer and Information Science and Engineering, University of Florida, USA}

% \linespread{0.9}

% \let\tempone\itemize
% \let\temptwo\enditemize
% \renewenvironment{itemize}{
% \tempone
%     \addtolength{\topsep}{-\baselineskip}
%     \addtolength{\partopsep}{-\baselineskip}
%     \addtolength{\itemsep}{-0.25\baselineskip}
% }{\temptwo}
% \let\tempone\enumerate
% \let\temptwo\endenumerate
% \renewenvironment{enumerate}{
% \tempone
%     \addtolength{\topsep}{-\baselineskip}
%     \addtolength{\partopsep}{-\baselineskip}
%     \addtolength{\itemsep}{-0.25\baselineskip}
% }{\temptwo}

% \columnsep{0.95}

% \renewenvironment{itemize}{\begin{itemize}
% \addtolength{\topsep}{-\baselineskip}
% \addtolength{\itemsep}{-\baselineskip}
% \end{itemize}}

% \setlist[itemize]{noitemsep, topsep=0pt}

%
% the rest is as usual
%

\begin{abstract}
Optimal recursive decomposition (or \dfn{DR-planning}) is crucial for analyzing, designing, solving or finding realizations of geometric constraint sytems. While the optimal DR-planning problem is NP-hard even for general 2D bar-joint constraint systems, we describe an \candrpcomplexity\ algorithm for a broad class of constraint systems that are isostatic or underconstrained. The algorithm achieves optimality by using the new notion of a canonical DR-plan that also meets various desirable, previously studied criteria. In addition, we leverage recent results on Cayley configuration spaces to show that the indecomposable systems---that are solved at the nodes of the optimal DR-plan by recombining solutions to child systems---can be minimally modified to become decomposable and have a small DR-plan, leading to efficient realization algorithms. We show formal connections to well-known problems such as completion of underconstrained systems.
Well suited
% Particularly amenable
to these methods are classes of constraint systems that can be used to efficiently model, design and analyze quasi-uniform (aperiodic) and self-similar, layered material structures.
% is the recursive decomposition of diverse types of underlying 2D geometric constraint systems, which we call \dfn{qusecs}.
%
We formally illustrate by modeling silica bilayers as body-hyperpin systems and cross-linking microfibrils as pinned line-incidence systems.
%
% We briefly discuss the modeling and analysis of two specific 2D layered materials, modeled as body-hyperpin and pinned line-incidence systems, as applications of the above theory and algorithms.
%
A software implementation of our algorithms and videos demonstrating the software are publicly available online\footnote{Visit \url{http://cise.ufl.edu/~tbaker/drp/index.html}.}.
\end{abstract}


\end{frontmatter}
% \maketitle
%%%%%%%%%%%%%%%%%%%%%%%%%%%%%%%%%%%%%%%%%%%%%%%%%%%%%%%%%%%%%%%%%%%%

% \input{s_response}

\section{Introduction}
\label{sec:intro}

\newcommand{\seedefs}{(formally defined in \ref{sec:appendix:defs})}
\newcommand{\seedefsb}{(see \ref{sec:appendix:defs} for definitions)}
\newcommand{\seedefsc}{See \ref{sec:appendix:defs} for formal definitions}
\newcommand{\seedefsd}{see \ref{sec:appendix:defs} for definitions}
\newcommand{\seedefsprelim}{(formally defined in Section \ref{sec:prelim:defs})}

% \todo{Regarding qusecs... We use the standard way of dealing w/ these problems, namely modeling it as a geometric constraint system...
% \todo{Restructure intro so that we have... problems first... closing with ``Concerning these issue... this is what we do in relation to them... the specifics we address..'' and go into summary of our contributions}

Geometric constraint systems have well-established, mature applications in mechanical engineering and  robotics, and they continue to find emerging applications in diverse fields from machine learning to molecular modeling.  Solving or realizing geometric constraint systems requires finding real solutions to a large multivariate polynomial system (of equalities and inequalities representing the constraints); this requires double exponential time in the number of variables, even if the type or orientation of the solution is specified.
% Thus, recursive decomposition into locally rigid subsystems (which have finitely many solutions) is crucial for realizing a geometric constraint system by the reverse process of recombining the subsystem solutions.
Thus, to realize a geometric constraint system, it is crucial to perform recursive decomposition into locally rigid subsystems (which have finitely many solutions), and then apply the reverse process of recombining the subsystem solutions.
With the use of \dfn{decomposition-recombination (DR-) planning}, the complexity is dominated by the size of the largest subsystem that is solved, or recombined, from the solutions of its child subsystems, i.e.\ the maximum fan-in occurring in a DR-plan.
In addition, navigating and analyzing the solution spaces, as well as designing constraint systems with desired solution spaces, leads to the \dfn{optimal decomposition-recombination (DR-) planning} problem~\cite{sitharam2005combinatorial, hoffman2001decompositionI, hoffman2001decompositionII}.

For a broad class of geometric constraint systems, local rigidity is characterized generically as a sparsity and tightness condition of the underlying constraint (hyper)graph~\cite{laman1970graphs,streinu2009sparse,tay1976rigidity,white1987algebraic}. This allows the generic DR-planning problem to be stated and treated as a combinatorial or (hyper)graph problem as we do in this paper.

We additionally conjecture that versions of \frontier, a previously given, bottom-up, polynomial time method for obtaining DR-plans with various desired properties~\cite{hoffman2001decompositionI, hoffman2001decompositionII,lomonosov2004graph,sitharam2005combinatorial}, in fact generate an optimal DR-plan for non-overconstrained systems.

Na{\"i}vely, the optimal DR-plan is used as follows. Each decomposed subsystem---a node of the DR-plan---is treated and solved as a polynomial system of constraints between its child subsystems. However, even in an optimal DR-plan, there can be arbitrarily many children at a node. In other words, even in the recursive decomposition given by an optimal DR-plan, the  size of the maximal indecomposable subsystem could be arbitrarily large.  It represents a bottleneck that dictates the complexity of solving or realizing the constraint system \cite{sitharam2010optimized,sitharam2006well,sitharam2010reconciling}.  We address this problem using the recently developed concept of \dfn{convex Cayley configuration spaces} \cite{sitharam2010convex,sitharam2011cayleyI,sitharam2011cayleyII,sitharam2014beast,sitharam2013caymos,wang2014cayley}. This allows for even greater reduction of the complexity by realizing large, indecomposable systems in a manner that avoids working with large systems of equations.
Specifically, we give an efficient technique for \dfn{optimally modifying} large indecomposable subsystems in a manner that reduces their complexity while preserving desired solutions; the modification ensures a convex Cayley configuration space, and the space can be efficiently searched to find a realization that satisfies the additional constraints of the original system.
This optimal modification  problem is a generalization of the previously studied problem of optimal completion of underconstrained systems \cite{sitharam2005combinatorial,joan-arinyo2003transforming}.

DR-plans are especially useful for constraint systems that exhibit some level of \dfn{self-similarity} and \dfn{quasi-uniformity}, in addition to isostaticity.  These properties can be leveraged to further reduce the complexity of both optimal DR-plan construction and recombination.
We consider 3 different types of constraint systems---which we collectively call \dfn{qusecs}---that are used to model, design, and analyze quasi-uniform or self-similar materials.
In the remainder of this section, we motivate the materials application and give the contributions and organization of the paper.

\subsection{Introducing Qusecs}
A large class of constraint systems that we call \dfn{qusecs}, a contraction of ``quasi-uniform or self-similar constraint system'', (a) can be treated combinatorially as described above and (b) occur as independent (isostatic or underconstrained) systems in materials applications. We discuss these next.
Some natural and engineered materials can be analyzed by treating them as two dimensional (2D) layers. As illustrated by the examples below, the structure within each layer is often: self-similar\footnote{In this manuscript we only study finite 2D structures. \dfn{Self-similarity} refers to the result of finitely many levels of hierarchy or subdivision in an iterated scheme to generate self-similar structures.}~\cite{2012arXiv1204.6389G}, spanning multiple scales; generally aperiodic and quasi-uniform within any one scale; and composed of a few repeated motifs appearing in disordered arrangements.
Note that a 2D layer is not necessarily planar (genus 0), it can consist of multiple, inter-constraining planar monolayers. Furthermore, a layer is often  either \dfn{isostatic} or \dfn{underconstrained} (not \dfn{self-stressed}/\dfn{overconstrained}, \seedefsd). These properties, as well as quasi-uniformity, aperiodicity, self-similarity, and layered structure, are natural consequences of evolutionary pressures or design objectives such as stability, minimizing mass, optimally distributing external stresses, and participating in the assembly of diverse and multifunctional, larger structures.

The importance of an optimal DR-plan is particularly evident for a qusecs. The quasi-uniform or self-similar properties mean that the decomposition and solution for one subsystem can be used as the decomposition and solution for other subsystems, thus causing further reduction in the complexity of both DR-planning and recombination. This is shown in Figures \ref{fig:c2c3ofk33s} and \ref{fig:bodypindrp}.

% \FigInit{}{fig:material_examples}
% \FigThreeSubfig%
%   {Chlamydomonas_TEM_17}
%   {Cross section of the Chlamydomonas algae axoneme, a cilia composed of microtubules~\cite{wikimediacommons2007cilia}.}
%   {fig:material_examples:microtubule}%
%   %
%   {ligten2}
%   {Cross section of a tendon displaying the hierarchical structure~\cite{lecture_biosolid_mechanics}.}
%   {fig:material_examples:tendon}%
%   %
%   {Rothemund-DNA-SierpinskiGasket}
%   {A DNA array exhibiting the Sierpinski triangle~\cite{wikimediacommons2007dna}.}
%   {fig:material_examples:sierpinski}

\ClearMyMinHeight
\SetMyMinHeight{.32}{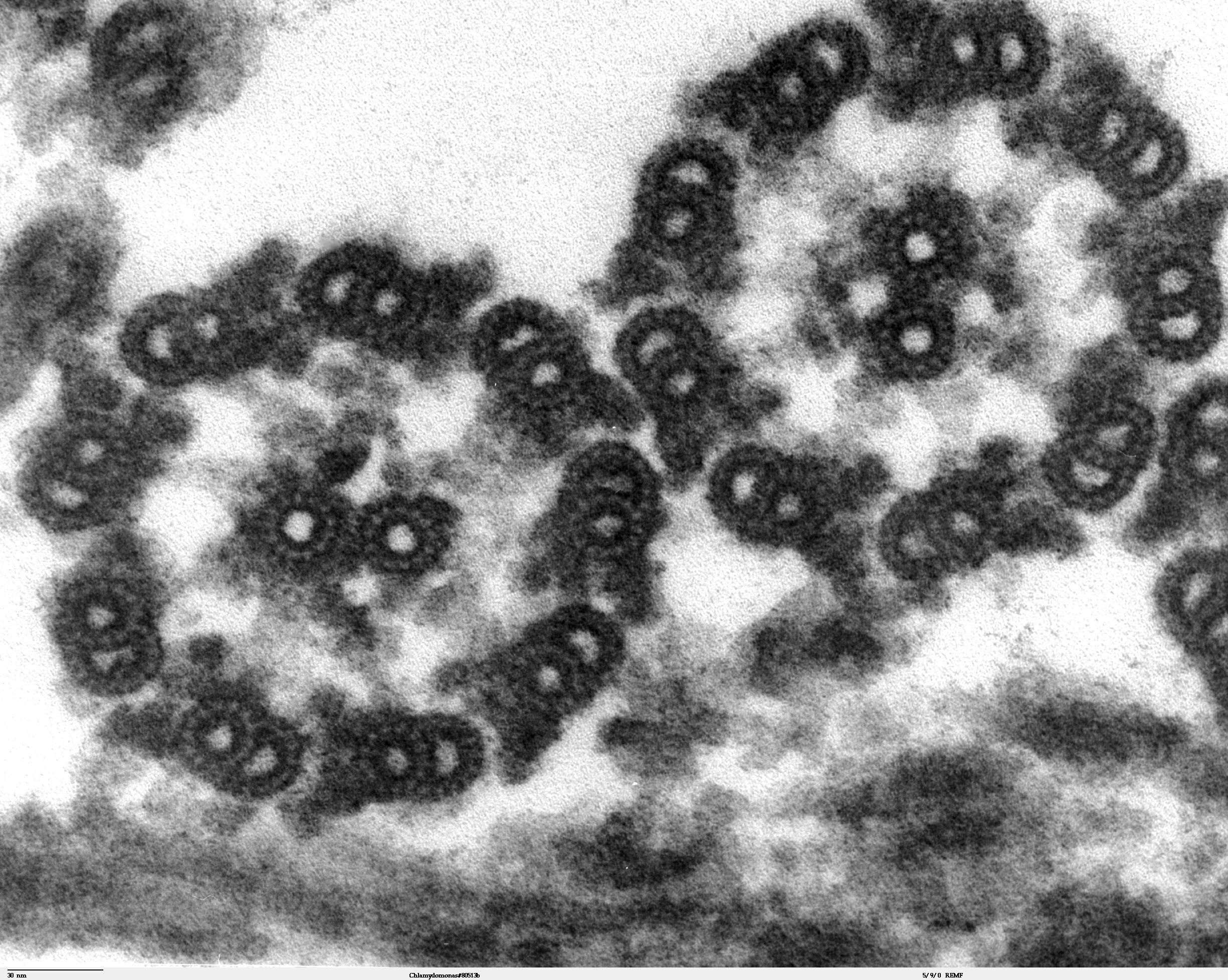}
\SetMyMinHeight{.32}{ligten2}
\SetMyMinHeight{.32}{Rothemund-DNA-SierpinskiGasket}

\begin{figure*}\centering%
  \begin{subfigure}{0.32\linewidth}\centering
    \includegraphics[height=\myMinHeight]{Chlamydomonas_TEM_17.jpg}
    \caption{}\label{fig:material_examples:microtubule}
  \end{subfigure}%
  \hfill
  \begin{subfigure}{0.32\linewidth}\centering
    \includegraphics[height=\myMinHeight]{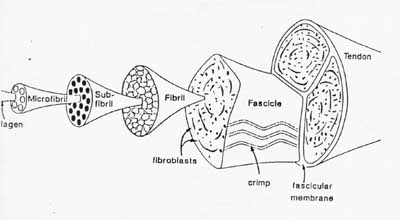}
    \caption{}\label{fig:material_examples:tendon}
  \end{subfigure}%
  \hfill
  \begin{subfigure}{0.32\linewidth}\centering
    \includegraphics[height=\myMinHeight]{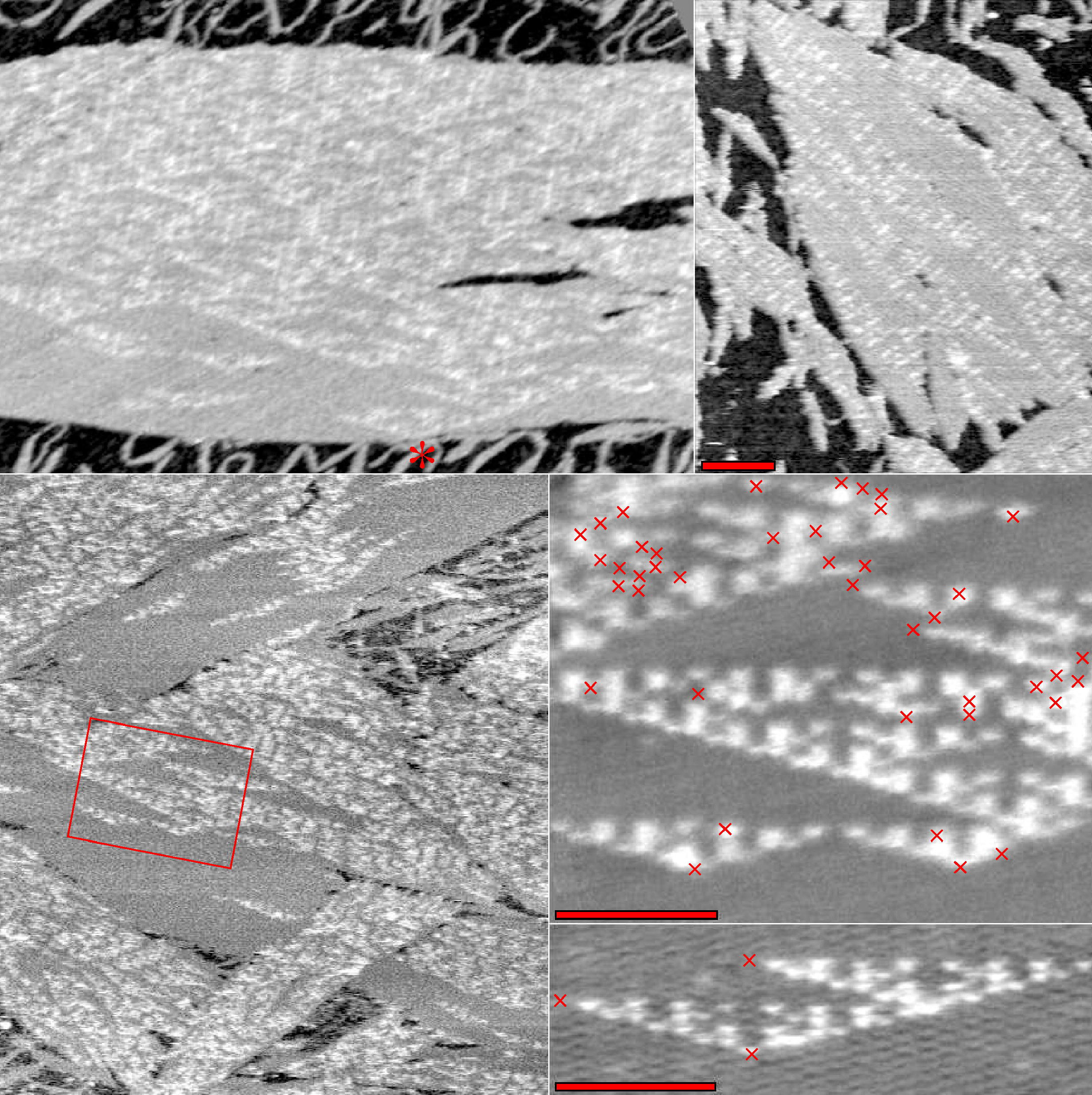}
    \caption{}\label{fig:material_examples:sierpinski}
  \end{subfigure}%
  \caption{(\ref{fig:material_examples:microtubule}) Cross section of the Chlamydomonas algae axoneme, a cilia composed of microtubules~\cite{wikimediacommons2007cilia}. (\ref{fig:material_examples:tendon}) Cross section of a tendon displaying the hierarchical structure~\cite{lecture_biosolid_mechanics}. (\ref{fig:material_examples:sierpinski}) A DNA array exhibiting the Sierpinski triangle~\cite{wikimediacommons2007dna}.}\label{fig:material_examples}
\end{figure*}%

% \FigAddSubfig%
%   {0.3}
%   {Chlamydomonas_TEM_17}
%   {Cross section of the Chlamydomonas algae axoneme, a cilia composed of microtubules~\cite{wikimediacommons2007cilia}.}
%   {fig:material_examples:microtubule}%
% \FigAddSubfig%
%   {0.3}
%   {ligten2}
%   {Cross section of a tendon displaying the hierarchical structure~\cite{lecture_biosolid_mechanics}.}
%   {fig:material_examples:tendon}%
% \FigAddSubfig%
%   {0.3}
%   {Rothemund-DNA-SierpinskiGasket}
%   {A DNA array exhibiting the Sierpinski triangle~\cite{wikimediacommons2007dna}.}
%   {fig:material_examples:sierpinski}
% \FigDisplay{}{fig:material_examples}

% Examples of such materials (See Figure \ref{fig:material_examples}) include:
% In order to study structural and mechanical properties of a material layer, it is natural to model a material layer as a solution or realization of a geometric constraint system of appropriate types of geometric primitives, under metric or algebraic constraints.
% Such 2D \dfn{qusecs}, quasi-uniform or self-similar constraint systems \seedefsprelim, can be used to understand or design material layers (their solutions) with desired properties.
Some materials that are readily modeled as qusecs include:
\begin{enumerate}
    \item \label{materialexample1} Cross-sections of microtubule structures~\cite{microtubule_necklace} (Figure \ref{fig:material_examples:microtubule}), e.g., in ciliary membranes and transitions~\cite{microtubule_cilia}.

    \item \label{materialexample2} Cross-sections of organic tissue with hierarchical structure, e.g., compact bone and tendon (Figure \ref{fig:material_examples:tendon}).

    \item \label{materialexample3} Crosslinked cellulose or collagen microfibril monolayers, e.g., in cell-walls~\cite{wikimediacommons2010afm, wikimediacommons2007plant}, as well as crosslinked actin filaments in the cytoskeleton matrix. See Section \ref{sec:pinnedline}.

    \item \label{materialexample4} More recent, engineered examples, including disordered graphene layers~\cite{Graphene1, Graphene2} sometimes reinforced by microfibrils; and DNA assemblies including a recent Sierpinski gasket~\cite{self_assembly_sierpinski} (Figure \ref{fig:material_examples:sierpinski}), bringing other self-similar structures~\cite{wikimediacommons2012subdivision} within reach.

    \item \label{materialexample5} Silica bi-layers~\cite{silica_bilayers}, glass~\cite{sructure_of_2d_glass}, and materials that behave like assemblies of 2D particles under non-overlap constraints, i.e.\ like jammed disks on the plane~\cite{jammed_disks}. See Section \ref{sec:bodypin}.
\end{enumerate}

\subsection{Organization and Contributions}
\label{sec:cont}

In Section \ref{sec:prelim}, we provide basic definitions in combinatorial rigidity theory,  and formalize the new notion of qusecs~\cite{sitharam2010optimized,sitharam2006well,sitharam2010reconciling}.
In addition, we define DR-plans and what it means for a DR-plan to be complete or optimal. We survey previous work on DR-planning algorithms, discussing other desirable criteria of DR-plans and their relation to the NP-hard optimality property of DR-plans.
%This section is relevant to Examples \ref{materialexample1} and \ref{materialexample2},  model them as bar-joint systems and discuss achieving isostaticity, distribution of stresses in self-similar, and other important concepts. \todo{We don't model them as bar-joint... What did this mean originally?}

% In Section \ref{sec:DRP}, we navigate the NP-hardness barrier (discussed in the following subsection), for finding optimal DR-plans by defining a so-called \dfn{canonical} DR-plan and showing a strong Church-Rosser property: \vemph{all canonical DR-plans for isostatic or underconstrained 2D qusecs are optimal}.

% Also in Section \ref{sec:DRP}, we give an efficient (\candrpcomplexity) algorithm to find a canonical (and hence optimal) DR-plan for all 3 types of 2D qusecs mentioned above (Sections \ref{sec:DRP}, \ref{sec:bodypin}, and \ref{sec:pinnedline}).
% The canonical DR-plan elucidates the essence of the NP-hardness of finding optimal DR-plans for over-constrained systems.
% Furthermore, our optimal/canonical DR-plan satisfies desirable properties such as the previously studied Cluster Minimality \cite{hoffman2001decompositionI} (see Figure~\ref{fig:demo_graph:clustmindrp}).

In Section \ref{sec:DRP}, we define a so-called \dfn{canonical DR-plan} and prove a strong Church-Rosser property: all canonical DR-plans for isostatic or underconstrained qusecs are optimal. In so doing, we navigate the NP-hardness barrier present in the general form of the DR-planning problem; the canonical DR-plan elucidates the essence of the NP-hardness of finding optimal DR-plans when a system is over-constrained. Furthermore, our optimal/canonical DR-plan satisfies desirable properties such as the previously studied \dfn{cluster minimality}~\cite{hoffman2001decompositionI}. Also in this section, a polynomial time (\candrpcomplexity) algorithm is provided to find a canonical DR-plan for isostatic bar-joint graphs.  While this and the next section focus on bar-joint graphs, the theory is easily extended to other qusecs used to model the abovementioned types of materials, as shown in subsequent sections.
% that such a plan will be optimal, in the sense that it minimizes the fan-in over all nodes, for isostatic bar-joint graphs.
% Additionally, a polynomial time (\candrpcomplexity) algorithm is provided to find a canonical DR-plan for isostatic bar-joint graphs.

In Section \ref{sec:recomb}, we give a method to deal with the algebraic complexity of recombining the realizations or solutions of child subsystems into a solution of the parent system \cite{sitharam2010optimized,sitharam2006well,sitharam2010reconciling}. Specifically, we define the problem of minimally modifying the indecomposable recombination system so that it becomes decomposable via a small DR-plan and yet preserves the original solutions in an efficiently searchable manner.
When the modifications are bounded, we obtain new, efficient algorithms for realizing both isostatic and underconstrained qusecs by leveraging recent results about Cayley parameters in \cite{sitharam2010convex,sitharam2011cayleyI,sitharam2011cayleyII} (see Sections \ref{sec:2-tree-reduction} and \ref{sec:tdecomp}).
In Section \ref{sec:table}, we show formal connection to well known problems such as optimal completion of underconstrained systems \cite{joan-arinyo2003transforming,sitharam2005combinatorial,gao2006ctree} and to find paths within the connected components.

% \todo{Section \ref{sec:recomb} addresses the issue of large, indecomposable subgraphs in the optimal DR-plan of bar-joint graphs by proposing a novel method of modification. By dropping edges to get a convex configuration space, realizations of the original linkage can be found. Also problem relations...}

 % The next sections focus on explicit applications by modeling certain materials as qusecs.

In Section~\ref{sec:bodypin} and~\ref{sec:pinnedline}, we briefly describe applications of the above techniques to modeling, analyzing, and designing specific properties in 2D material layers~\cite{Jackson2008bodypin}. We explicitly model these materials as qusecs. For Examples~\ref{materialexample4} and~\ref{materialexample5}, we discuss boundary conditions for achieving various desired properties of body-hyperpin systems. For Example~\ref{materialexample3}, we discuss canonical and optimal DR-plans for pinned line incidence systems~\cite{sitharam2014incidence}.

The last Section~\ref{sec:conclusion} concludes the paper, and Section \ref{sec:open} lists open problems and conjectures. In particular, we conjecture that the methods of Section~\ref{sec:DRP} extend in fact to a large class of (hyper)graphs, formally those with an underlying abstract rigidity matroid in which independence corresponds to some type of sparsity, and maximal independence (rigidity) is a tightness condition.

Throughout this paper,  an asterisk after a formal statement indicates that its proof appears in~\ref{sec:appendix:proofs}.

A software implementation of our algorithms and videos demonstrating the software are publicly available online\footnote{See footnote 1.}.

\section{Preliminaries and Background}
\label{sec:prelim}

% Basics definitions and theory for this topic can be found in~\ref{sec:appendix:defs}.

We first give basic definitions and concepts in combinatorial rigidity, leading to a definition of a DR-plan, its properties, and how they relate. The section ends with a discussion of previous work on DR-plans.

\subsection{Geometric Constraint Systems and Combinatorial Rigidity}
\label{sec:prelim:defs}
\label{sec:appendix:defs}

In this paper, a \dfn{geometric constraint system} is a multivariate polynomial (usually bilinear or quadratic) system $G(x,\delta)=0$, representing constraints with parameters $\delta$ between geometric primitives  in $\mathbb{R}^2$ represented collectively as $x\in \mathbb{R}^n$.
When the type of constraint (system) is fixed, the system is simply represented as $(G,\delta)$, where $G$ is the underlying constraint (hyper)graph $G = (V,E)$ with the vertices $V$ representing the geometric primitives in $\mathbb{R}^2$ and (hyper)edges $E$ representing the constraints, each with an associated parameter $\delta$.
For example, a \dfn{bar-joint system or linkage} $(G,\delta)$, is a graph $G=(V,E)$ with fixed length bars as edges,
% \footnote{Geometric constraint systems can also have inequalities in addition to equations, where the parameters in $\delta$ are small intervals of values rather than exact values.}
i.e.\ $\delta: E \rightarrow \mathbb{R}$; this represents the distance constraint system $\| x_u -x_v \|_2 = \delta_{u,v}$ for  $(u,v) \in E$, where $x_u \in \mathbb{R}^2$ represents the coordinates of $u\in V$.

In all types of geometric constraint systems we consider in this paper, a Cartesian \dfn{realization} or \dfn{solution} $G(p)$ of $(G,\delta)$ is an assignment of coordinates or Euclidean transformations (poses), $p: V \rightarrow \mathbb{R}^2$ or $\mathbb{R}^3$, to the vertices of $G$ satisfying the constraints with parameters $\delta$, modulo orientation preserving isometries (Euclidean rigid body motions).

Although the realization space itself depends on the constraint parameters $\delta$, many relevant \dfn{generic} properties of the constraint system $G(x,\delta)$ are defined to be properties of the constraint (hyper)graph $G$ and do not depend on $\delta$ (or they hold for all but a measure zero set of $\delta$ values). Many of these are properties of the Jacobian $\Delta_x G(x,\delta)$, often called the appropriate \dfn{rigidity matrix of $G$} (a matrix of indeterminates). For example, the \dfn{bar-joint rigidity matrix of the graph $G = (V,E)$} is a matrix of indeterminates representing the Jacobian of the distance map $\| x_u -x_v \|_2$  for $(u,v) \in E$. The matrix has $2$ columns per vertex in $V$ and one row per edge in $E$, where the row corresponding to edge $(u,v)$ contains the 2 coordinate indeterminates for $x_u -x_v$ (resp. $x_v-x_u$) in the 2 columns for $u$ (resp.\ $v$), i.e.\ 4 non-zero entries per row.

% XXXXXXXXX
% something has to be said along with asimow roth
% (infinitesimal rigidity implies rigidity) When the rigidity matrix has
% appropriate rank, the realizations or solutions of the corresponding
% constraint system are generically isolated and zero-dimensional if the
% constraint system has equalities and exact values for the parameters
% $\delta$. If the constraint system has inequalities, i.e, if the
% $\delta$ lie in an interval, this claim is approximate: the solutions
% are isolated small, full-dimensional  connected components.

One important property of a generic constraint system or (hyper)graph\footnote{We refer to these as properties of the constraint system or as properties of the underlying (hyper)graph interchangeably}
%that depends on the appropriate rigidity matrix of indeterminates
is \dfn{rigidity},
% (\note footnote now...),
i.e.\ the realizations or solutions of the corresponding constraint system being generically isolated and zero-dimensional.
%(if the constraint system has equalities and exact values for the parameters $\delta$).
The result by Asimow and Roth \cite{asimow1978rigidity}  shows a constraint (hyper)graph is rigid if and only if  it is generically \dfn{infinitesimally rigid}, i.e.\ the number of independent rows of its appropriate rigidity matrix is at least the number of columns less the number of rigid body motions, which is 3 for 2D bar-joint systems.
%This condition is called generic {\em infinitesimal rigidity}.

% \sidenote{If the constraint system has inequalities, i.e, if the $\delta$ lie in an interval, the definition of rigidity is approximate: the solutions are isolated small, full-dimensional  connected components.}
% generic rigidity, i.e.\ there exist at most finitely many solutions to the system at generic points,  %to the algebraic system

Geometric constraint systems can also have inequalities in addition to equations, where the parameters in $\delta$ are small intervals rather than exact values. In this case, the definition of rigidity is approximate; the solutions are isolated, small, full-dimensional connected components.

Other generic constraint system or (hyper)graph properties are mentioned here.
% (\note we refer to these as properties of the constraint
% system or as properties of the underlying (hyper)graph
% interchangeably).
A constraint (hyper)graph $G$ is \dfn{independent} if its appropriate rigidity matrix of indeterminates has independent rows (i.e.\ the determinant of some square submatrix is not identically zero).
%It is {\em rigid} if the number of independent rows of the rigidity matrix is
%at least the number of columns less the number of rigid body motions,
%which is 3 for distance constraint systems.
It is \dfn{isostatic} (\dfn{minimally rigid}, \dfn{wellconstrained}) if it is both rigid and independent.
%if the number of generically independent rows or the rank of the appropriate rigidity matrix is maximal.
%For example, the maximal rank of a bar-joint rigidity matrix is $2|V| - 3$,
%where $3$ is the number of rotational and translational degrees-of-freedom of a rigid body in $\mathbb{R}^2$.
%The graph $G$ is {\em rigid} if there exists some spanning subgraph $S\subseteq G$ such that $S$ is wellconstrained.
It is \dfn{flexible} if it is not rigid, \dfn{underconstrained} if it is independent and not rigid, or \dfn{overconstrained} if it is not independent.

Defining the combinatorial independence of a subset of edges $E'\subseteq E$ to be the independence of corresponding rows in the rigidity matrix of indeterminates, we obtain the \dfn{rigidity matroid} of a constraint (hyper)graph $G = (V,E)$.
%The 2-dimensional {\em rigidity matroid} of a constraint (hyper)graph $G = (V,E)$ is a linear matroid  on $E$,
%where a subset of edges $E' \subseteq E$ is {\em independent} in the matroid,
%if the set of corresponding rows in the appropriate rigidity matrix of indeterminates are linearly independent.
There are various results on combinatorial characterization of independence, rigidity, and rigidity matroids for different types of (hyper)graphs. For bar-joint rigidity matroids, the famous Laman's theorem \cite{laman1970graphs} states that the underlying graph is isostatic if and only if $|E| = 2|V|-3$ and $|E'| \le 2|V'|-3$ for every induced subgraph with at least 2 vertices. The result by Lovasz and Yemini \cite{lovasz1982generic} shows that all 6-vertex-connected graphs are rigid in the plane. For bar-body rigidity matroids, Tay \cite{tay1976rigidity} proved that the underlying multigraph is isostatic if and only if it can be decomposed as $3$ edge disjoint spanning trees. White and Whiteley \cite{white1987algebraic} gave the same characterization using a different technique to study the algebraic-geometric conditions of genericity, called pure condition. Lee, Streinu and Theran \cite{lee2007graded} defined the \dfn{$(k,l)$-sparsity matroid}, where a hypergraph $G$ is called \dfn{$(k,l)$-sparse} if $|E'| \le k|V'| - l$ for any induced subgraph $(V',E')$ with at least 2 vertices, and \dfn{$(k,l)$-tight} if it is $(k,l)$-sparse and $|E| = k|V| - l$. In general, given a $d$-uniform hypergraph, a $(k,l)$-sparsity condition is matroidal as long as $l \le dk-1$.

In this paper, a \dfn{qusecs} is any independent geometric constraint system of one of 3 types: bar-joint (defined formally in \ref{sec:prelim:defs}), \dfn{body-hyperpin} (defined formally in Section \ref{sec:bodypin}), and \dfn{pinned line-incidence} (defined formally in Section \ref{sec:pinnedline}).

We note that  the remainder of this section and Sections~\ref{sec:DRP} and \ref{sec:recomb} we only consider bar-joint qusecs and graphs. Relevant formal analogies for the other 2 types of qusecs and (hyper)graphs are given along with their materials applications in Sections \ref{sec:bodypin} and \ref{sec:pinnedline}.

\subsection{Decomposition-Recombination (DR-) Plans}

\ClearMyMinHeight
\SetMyMinHeight{.4}{3xc2c3}
\SetMyMinHeight{.3}{3xc2c3_comdrp}
\SetMyMinHeight{.3}{3xc2c3_candrp}

\begin{figure*}\centering%
  \begin{subfigure}{0.4\linewidth}\centering
    \includegraphics[height=\myMinHeight]{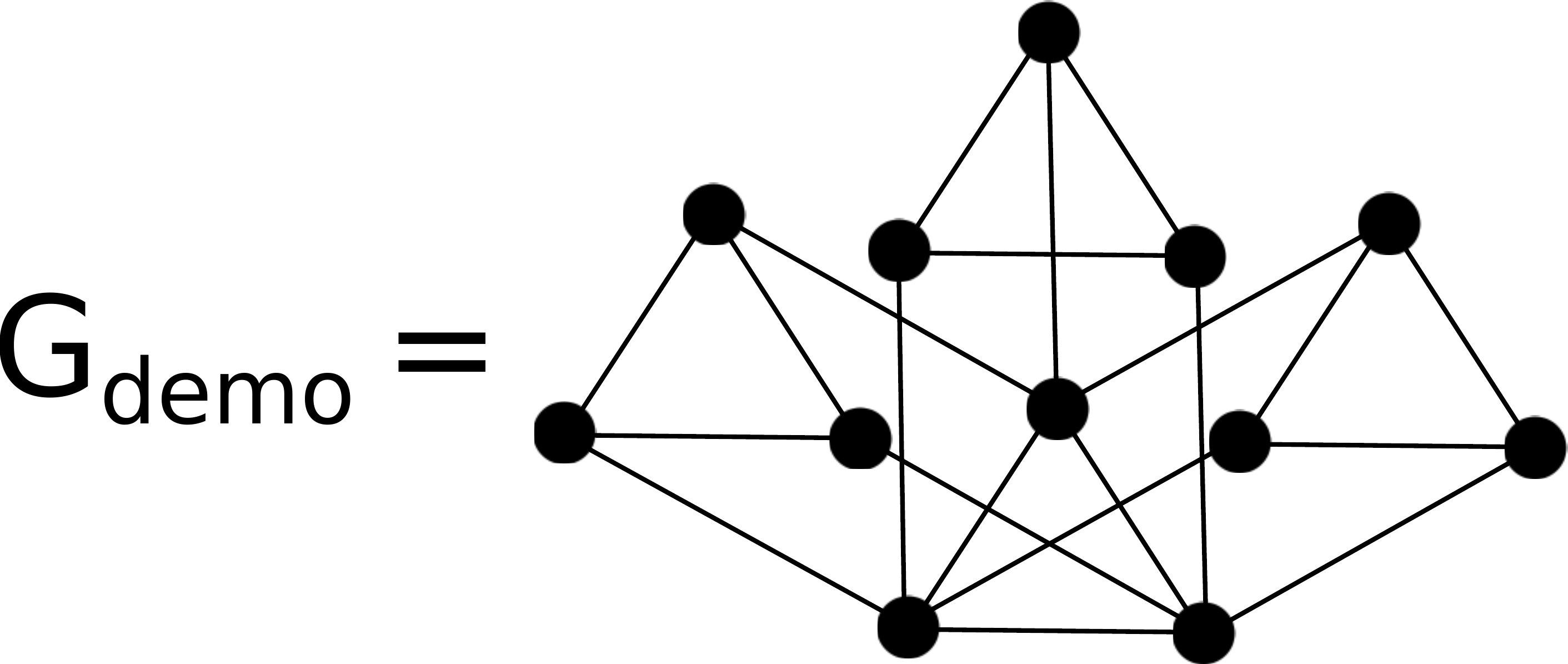}
    \caption{}\label{fig:demo_graph:graph}
  \end{subfigure}%
  \hfill
  \begin{subfigure}{0.3\linewidth}\centering
    \includegraphics[height=\myMinHeight]{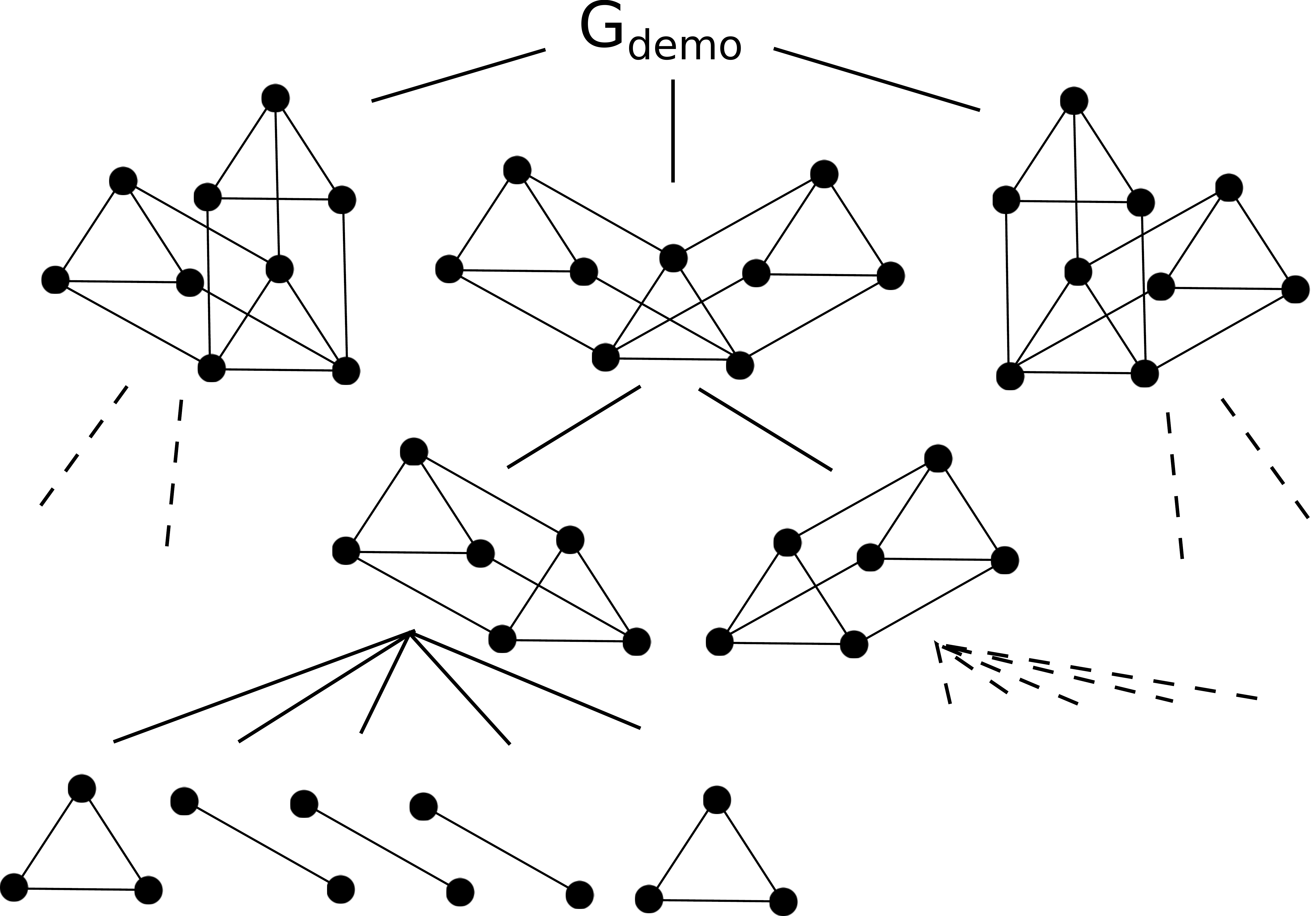}
    \caption{}\label{fig:demo_graph:comdrp}
  \end{subfigure}%
  \hfill
  \begin{subfigure}{0.3\linewidth}\centering
    \includegraphics[height=\myMinHeight]{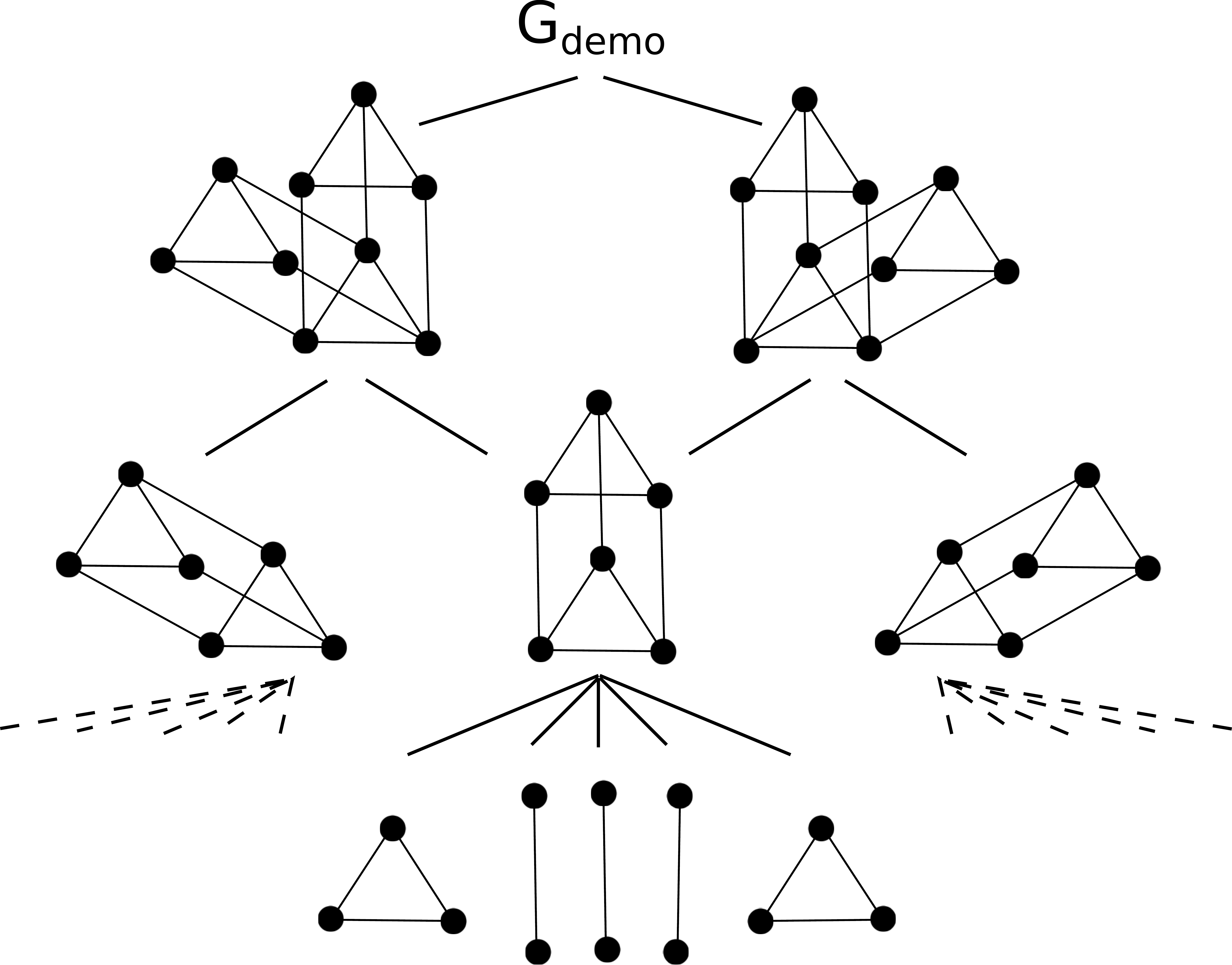}
    \caption{}\label{fig:demo_graph:candrp}
  \end{subfigure}%
  \caption{(\ref{fig:demo_graph:graph}) A graph, $G_{demo}$, used to illustrate concepts throughout this and the next section. (\ref{fig:demo_graph:comdrp}) The complete DR-plan of $G_{demo}$, i.e.\ $ComDRP(G_{demo})$. Dashed lines indicate that the children repeat the same pattern as the others shown on this level. The children of triangles (3 edges) are omitted. (\ref{fig:demo_graph:candrp}) The canonical DR-plan of $G_{demo}$, which is optimal (see Section~\ref{sec:DRP}), i.e.\ $OptDRP(G_{demo})$. The children of triangles are omitted.}
  \label{fig:demo_graph}
\end{figure*}%

% \ClearMyMinHeight
% \SetMyMinHeight{.4}{triangle}
% \SetMyMinHeight{.3}{triangle_candrp}
% \SetMyMinHeight{.3}{triangle_clustmindrp}

%\begin{figure*}\centering%
%
%\begin{subfigure}{0.4\linewidth}\centering
%\includegraphics[height=\myMinHeight]{triangle}
%    \caption{}\label{fig:demo_graph_tri:graph}
% \end{subfigure}%
%
%  \hfill
% \begin{subfigure}{0.3\linewidth}\centering
%   \includegraphics[height=\myMinHeight]{triangle_candrp}
%    \caption{}\label{fig:demo_graph_tri:candrp}
%  \end{subfigure}%
%
%  \hfill
%  \begin{subfigure}{0.3\linewidth}\centering
%   \includegraphics[height=\myMinHeight]{triangle_clustmindrp}
%   \caption{}\label{fig:demo_graph_tri:clustmindrp}
%  \end{subfigure}%
%
%  \caption{(\ref{fig:demo_graph_tri:graph}) A simple graph, $G_{tri}$, of 3 triangles intersecting trivially. (\ref{fig:demo_graph_tri:candrp}) The canonical DR-plan of $G_{tri}$. (\ref{fig:demo_graph_tri:clustmindrp}) A DR-plan of graph $G_{tri}$ which gives cluster minimality  (See Section \ref{sec:prev}). Using the ideas from the paper~\cite{lomonosov2004graph}, the children of a triangle are the primitives of a point and an edge. Thus, the fan-in of this graph is 3, whereas the canonical/optimal DR-plan has fan-in of 2. With this counterexample, it is clear that cluster minimality is not sufficient for optimality.}
% \label{fig:demo_graph:clustmindrp}
%  \label{fig:demo_graph_tri}
%\end{figure*}%

\begin{definition}\label{def:drp}
    The \dfn{decomposition-recombination \mbox{(DR-)} plan} \cite{hoffman2001decompositionI} of graph $G$, $\drp{G}$, is defined as a forest that has the following properties:
    \begin{enumerate}
        \item Each node represents/contains/is a rigid subgraph of $G$.
        % \item For a node, $C$, that is  a non-trivial graph, its $N$
        % children, $C_1, \ldots, C_N$, are rigid vertex-maximal proper
        % subgraphs of $C$.
        \item The children $C_1,\ldots,C_N$ of a node $C$ satisfy $\bigcup_{i=1}^N{C_i}=C$.
        \item A leaf node is a single edge. A \dfn{trivial} graph is empty or a single vertex. Note that a trivial graph is not isostatic.
        \item A root node is a vertex-maximal rigid subgraph of $G$.
    \end{enumerate}
%
    % It can also be described recursively as: the root is $G$, its children are the trivial subgraphs and the DR-plans of its wellconstrained vertex-maximal proper subgraphs whose union is $G$ itself.
%
    A DR-plan is \dfn{complete} if it satisfies an additional property: for a non-leaf node $C$, its children are all of the rigid vertex-maximal proper subgraphs of $C$. This makes Property 2 implicit.
  %and in fact, $C$ contains all the edges in the union of its children as well.
  We denote a complete DR-plan of $G$ as $\comdrp{G}$.

  A DR-plan is \dfn{optimal} if it minimizes the maximum fan-in over all nodes in the tree. The maximum  fan-in is called the \dfn{size} of the DR-plan. We denote an optimal DR-plan of $G$ as $\optdrp{G}$.
%
%In general, $C$ is the graph in an arbitrary node in $CompleteDRP(G)$.
%$C_i$ is the $i^{\text{th}}$ child of $C$. It is implied that $C_i$ is
%an isostatic vertex-maximal proper subgraph.
    %Note that nodes will be referred to interchangeably as ``the node that
    %represents or contains the (sub)graph $C$'' and as simply ``$C$''.
\end{definition}
%
% \todo{This was labeled as both begin(remark) and end(observation), what was your intent?}
\begin{remark}
More than one node (leaf) in a DR-plan forest may represent the same subgraph (vertex) of $G$.
For a given graph, there could be exponentially many DR-plans---and even optimal DR-plans---in the size of the graph. A complete DR-plan is unique but may not be (and is usually not) optimal. DR-plans of self-similar graphs are self-similar.
\end{remark}

See Figures \ref{fig:demo_graph}, \ref{fig:c2c3ofk33s}, \ref{fig:demo_graph:candrpseq}, \ref{fig:overconstrained}, and \ref{fig:bodypindrp} for examples of DR-plans and how their properties relate to each other.

\subsection{Previous Work on DR-Plans}
\label{sec:prev}
% \todo{We now briefly survey existing techniques for studying 2D qusecs,  many of which are \dfn{bar-joint} systems (Examples 1 and 2 above, see Sections \ref{sec:prelim}, \ref{sec:DRP}, and \ref{sec:recomb}), \dfn{body-hyperpin} systems (Example 4 and 5, see Section \ref{sec:bodypin}) or \dfn{pinned-line incidence} systems (Example 3, see Section \ref{sec:pinnedline}). The limitations of these techniques directly motivate the contributions of this paper.}
We now briefly survey existing techniques for detecting rigidity and creating DR-plans of 2D constraint systems. The limitations of these techniques directly motivate the contributions of the next  section.

\subsubsection{Finding (Vertex)-Maximal, Generically Rigid Subsystems}
Fast, graph-based algorithms exist (pebble-game~\cite{Jacobs:1997:PG,hoffmann1997solvablesubsets,jermann2006decomposition,Lee:2007:PGA}), for locating all maximal, \dfn{generically rigid} subsystems \seedefs. When the input itself is rigid, these algorithms do nothing, i.e.\ compute the identity function.

However, both for self-similar or just aperiodic 2D qusecs, it is imperative to recursively decompose rigid systems into their rigid subsystems, down to the level of geometric primitives, in order to understand or design properties at all scales, such as \seedefs\ \dfn{rigidity}, \dfn{flexes}, distribution of \dfn{external stresses}, boundary conditions for \dfn{isostaticity}, as well as behavior under constraint variations.

\subsubsection{Optimal Recursive Decomposition (DR-Planning)}
Recursive decomposition of geometric constraint systems has been formalized \cite{hoffman2001decompositionI,hoffman2001decompositionII} and well-studied \cite{lomonosov2004graph,sitharam2005combinatorial,jermann2006decomposition} as the \dfn {Decomposition-Recombination (DR-) planning} problem \seedefsprelim. For the abovementioned classes of 2D qusecs, generic rigidity is a combinatorial property and hence each level of the decomposition should, in principle, be achievable by a graph-based algorithm without involving the geometric information in the constraint system. Since many such decompositions can exist for a given constraint system, criteria defining desirable or optimal DR-plans and DR-planning algorithms were given in~\cite{hoffman2001decompositionI}.
%An \dfn{optimal DR-plan} is one that minimizes the \dfn{size} \seedefsprelim, i.e., the maximum number of child subsystems of any parent system. Being exponential in the size, the complexity of solving the parent constraint system is overwhelmingly dominated by the complexity of solving the child systems.

However, for overconstrained 2D qusecs, even when restricted to bar-joint systems, the optimal DR-planning problem was shown to be NP-hard \cite{lomonosov2004graph, sitharam2005combinatorial}.
The NP-hardness of the optimal DR-planning problem for 2D bar-joint graphs is partly the consequence of possibly exponential number of DR-plans. On the other hand, although the complete DR-plan is unique it could have large average fan-in and exponentially many nodes making it far from optimal.

\subsubsection{DR-plans for Special Classes and with Other Criteria}
For a special class of 2D qusecs, namely \dfn{tree-decomposable} systems~\cite{owen1991algebraic,fudos1997graph,joan-arinyo2004revisiting} common in computer aided mechanical design (which includes ruler-and-compass and Henneberg-I constructible systems), all DR-plans turn out to be optimal. This satisfies the Church-Rosser property, leading to highly efficient DR-planning algorithms. For general 2D qusecs, alternate criteria were suggested such as \dfn{cluster minimality} requiring parent systems to have a minimal set of at least 2 rigid proper subsystems as children (i.e.\ the union of no proper subset of size at least 2 child subsystems forms a rigid system); and \dfn{proper maximality}, requiring child subsystems to be maximal rigid proper subsystems of the parent system. \seedefsc.

While polynomial time algorithms were given to generate DR-plans meeting the cluster minimality criterion~\cite{lomonosov2004graph}, no such algorithm is known for the latter criterion.

\section{Main Result: Canonical DR-Plan, Optimality, and Algorithm}
\label{sec:DRP}
% In the problem of the optimal DR-Plan there is generally not a unique plan.
% % Indeed, we will prove a union of $N$ isostatic subgraphs will result in $N$ unique plans, but that at the $N^{\text{th}}$ level of the tree it will always be the same. Therefore, all choices of decomposition are in some sense equivalent. The theorem we seek to prove is thus:
% However, we will show that regardless of which children are chosen for the plan, so long as they satisfy the definition of an optimal DR-plan, the recombination will require solving of the same systems. Being the smallest such structure that offers this, the definition of an optimal DR-plan could be considered the canonical DR-plan.
% % To assist in showing this, we prove this core theorem throughout this section:

% In this section, we discuss 2D bar-joint graphs. All vertex weights are $2$, all edge weights are $1$, and constant $k= -{{3}\choose{2}}=-3$. Trivial graphs are a single vertex and empty set. Furthermore, 2D isostatic graphs must be connected.
% % The greatest density of a 2D isostatic graph is $-2$ (the vertex). The other disconnected part of the graph would need to have a density of $-1$, which is overconstrained and not possible in a isostatic graph (because there is no trivial graph with that density).

\subsection{Canonical DR-Plan}
In this section, we define a \dfn{canonical} plan to capture those aspects of an optimal DR-plan that mimic the  uniqueness of a complete DR-plan, and we show that the nonunique aspects do not affect optimality for independent (underconstrained or isostatic) graphs. Furthermore, we give an efficient \candrpcomplexity\ algorithm to find the canonical DR-plan of any independent graph. The definition is as follows:

\begin{definition}\label{def:canonical_drplan}
    The \dfn{canonical DR-plan} of a graph $G$ satisfies the following three properties:
    (1) it is a DR-plan of $G$;
    (2) children are rigid vertex-maximal proper subgraphs of the parent; and
    (3) if all pairs of rigid vertex-maximal proper subgraphs intersect trivially then all of them are children, otherwise exactly two that intersect non-trivially are children.
\end{definition}

In this section and in section~\ref{sec:recomb}, any reference to a graph $G$ is assumed to be isostatic (i.e.\ well-constrained or $(k,l)$-tight).
% \sidenote{In this section, $G$ is always assumed to be isostatic.}

% This definition gives the canonical DR-plan a surprisingly strong Church-Rosser property.

Definition~\ref{def:canonical_drplan} gives the canonical DR-plan a surprisingly strong Church-Rosser property, which is made explicit in Theorem~\ref{theorem:main}, the main result of this section.

\begin{theorem}
\label{theorem:canonical_exists_and_is_optimal}
\label{theorem:canonical_is_optimal}
\label{theorem:main}
    A canonical DR-plan exists for a graph $G$ and any canonical DR-plan is optimal if $G$ is independent.
\end{theorem}

% \begin{theorem} \label{theorem:canonical_is_optimal}
%     \label{theorem:main}
%     Any canonical DR-plan is an optimal DR-plan.
% \end{theorem}

%The proof of this theorem is a direct consequence of the following  more general theorem.

%\begin{theorem}\label{theorem:main}
%Given an isostatic 2D bar-joint graph $G$ and ComDRP$(G)$, for all nodes $C$
%with children $C_1,\ldots,C_N$ preserve children according to the following rules.
%\begin{enumerate}
%    \item If $C_i \cap C_j$ is trivial then keep all $C_1,\ldots,C_N$ as children.
%    \item If $C_i \cap C_j$ is isostatic then select any two out of $C_1,\ldots,C_N$ as children.
%\end{enumerate}
%This is a canonical DR-plan.
%\end{theorem}

\begin{proof}
We show the existence of a canonical DR-plan by constructing it as follows:

Begin with $\comdrp{G}$ of a rigid 2D bar-joint graph $G$, for all nodes $C$ with children $C_1,\ldots,C_N$  retain children nodes according to the following rules:
\begin{enumerate}
   \item If $C_i \cap C_j$ is trivial then retain all $C_1,\ldots,C_N$ as children.
   \item If $C_i \cap C_j$ is rigid then select any two out of $C_1,\ldots,C_N$ as children.
\end{enumerate}

This directly satisfies Properties (2) and (3) of a canonical DR-plan (see Definition~\ref{def:canonical_drplan}), because all the nodes in $\comdrp{G}$ are rigid vertex-maximal proper subgraphs,  which we shorten to {\em clusters}.  To show Property (1) holds (that this constitutes a DR-plan):
for Case 1 above,  since we start with a complete DR-plan, if we preserve all the children it is still a DR-plan; for Case 2 above, we know that the union must be rigid as well and it cannot be anything other than $C$, otherwise we would have found a larger rigid proper subgraph of $C$, contradicting vertex-maximality.

Note that if we begin with an isostatic graph, ``rigid'' can be replaced with ``isostatic'' throughout the construction and preserve the above properties. The rigid proper subgraphs of an isostatic graph must be isostatic themselves.

Next we show that a canonical DR-plan is optimal.

First, note  that any  DR-plan $R$ without the Property (2) of a canonical DR-plan can always be modified (by introducing intermediate nodes) to satisfy Property (2) without  increasing the max fan-in, since any rigid proper subgraph of a graph $C$ (a child of node $C$ of the DR-plan $R$)
is the subgraph of some cluster of $C$.
Thus without loss of generality, we can assume that an optimal DR-plan satisfies Property (2) of a canonical DR-plan.

The proof of optimality of a canonical DR-plan is by induction on its height.  The base case trivially holds for canonical DR-plans of height 0, i.e.\ for single edges. The induction hypothesis is that canonical DR-plans of height $t$ are optimal for the root node.
For the induction step consider a canonical DR-plan $R$ of  height $t+1$ rooted at a node $C$.
Notice that $R$ represents a canonical DR-plan $R(C)$ for the graphs $C$ corresponding to each of its  descendant nodes.
Thus, from the induction hypothesis, we know that the $R(C_i)$ is optimal for $C_i$.

Thus it is sufficient to demonstrate a set of nodes $S$  that must be present in any  DR-plan $R$ for $C$ that satisfies Property (2), including a known optimal one; and furthermore, for any such DR-plan $R$,  either (Claim 1) $S$ must be the set of children of $C$; or (Claim 2) for all the ancestors $A$ of $S$, $R$ has the minimum possible fan-in of 2.

We show the two claims below.
The first claim is that for a node $C$ whose clusters have trivial pairwise intersections, any DR-plan of $C$ that satisfies Property (2) must also satisfy Property (3) at $C$, i.e.\ the set of children $S$ of $C$ consists of all clusters of $C$.
Because this is the only choice, it is the minimum fan-in at $C$ for any DR-plan for $C$ with Property (2), including a known optimal one.
The second claim shows that in the case of nodes $C$ whose rigid, vertex-maximal proper subgraphs have  non-trivial pairwise intersections, every canonical DR-plan of $C$ that uses any possible choice of two such subgraphs of $C$ as children results in a minimum possible fan-in of 2 in the ancestor nodes $A$ leading to the {\em same maximal antichain $S$ of descendants $D$ of $C$}. The antichain is maximal in the partial order of rigid subgraphs of $C$ under containment. I.e.\ $S$ satisfies the property that every proper vertex-maximal rigid subgraph of $C$ is a superset of some $D$ in $S$; this follows from properties of maximal antichains that no element of $S$ is contained in the union of other elements of $S$; and the union of elements of $S$ is $C$. Thus any  DR-plan that satisfies Property (2) and hence contains two or more of the rigid vertex-maximal proper subgraphs of $C$ as children must also contain every element of $S$. The two claims complete the proof that every canonical DR-plan is optimal.

 % We do this inductively on the l, beginning with the root node, showing that the rules are always the optimal choice.
% by discussing each rule and proving inductively that this is the optimal choice at each level of the DR-plan tree, starting with the root node.

% Now we show that this plan is optimal by considering the two cases. Observation \ref{lemma:union_intersection} shows that we do not have to consider anything other than the two cases stated in the construction.
 % the intersection of any two isostatic subgraphs can only result in trivial or isostatic subgraphs. Therefore, given $C$ and its isostatic vertex-maximal subgraphs $C_1,\ldots,C_N$, the are only two possibilities to consider. Either (1) subgraphs $C_i$ and $C_j$ have a trivial intersection, or (2) they have an isostatic intersection.

\medskip\noindent
\vemph{Claim 1:}
% If some pair is intersecting trivially, then in fact all pairs intersect trivially, thus all must be children by defn of canonical drplan
%
A set of clusters $C_1,\ldots,C_N$ whose pairwise intersection is trivial, must be children of $C$ in an optimal DR-plan.

We prove this claim  by showing that the union of no subset of the children can be $C$, thereby requiring all of them to be included as children.

% must be optimal for this node.

We prove by contradiction.
Assume to the contrary that the strict subset $S\subsetneq \{1,\ldots,N\}$ such that $U=\bigcup_{i\in S}{C_i}$ is isostatic. If $U\neq C$, then we found a larger proper subgraph contradicting vertex-maximality of the $C_i$. So, it must be that $U=C$.
\usestwod
However, since $C_i \cap C_j$ is trivial then for $k\notin S$ we know, by Lemma \ref{lemma:combined_lemma}, Item \ref{lemma:uc_intersection_makes_all_uc}, $U\cap C_k$ must be one or more trivial, i.e.\ disconnected vertices. By definition of a DR-plan, $C_k=C\cap C_k$ and we know that $U=C$ so $C_k=U\cap C_k$. Thus, $C_k$ is (i) a collection of disconnected vertices, and (ii) an isostatic subgraph of $C$, which is impossible. As $C$ is isostatic, this means the union of no proper subset of $C_1,\ldots,C_N$ is isostatic, nor is it equal to $C$, proving Claim 1.

Furthermore, since a canonical DR-plan has nodes with proper rigid \vemph{vertex-maximal} subgraphs as children, if, as in this case, their pairwise intersection is trivial, it follows that any node has at most as many children as a DR-plan without this restriction, because the union of the children must contain all edges of the parent. Therefore, the canonical DR-plan is the optimal choice in this case of trivial intersections.

\medskip\noindent
\vemph{Claim 2:}
If some pair in the set of child clusters $C_1,\ldots,C_N$ of $C$ has an isostatic (nontrivial) intersection, then choosing any two as children (minimum possible fan-in) will result in the same maximal antichain of descendants of $C$.

To prove Claim 2, notice that if  $C_i \cap C_j$ is isostatic, then, by Observation \ref{lemma:union_intersection}, $C_i \cup C_j$ is also isostatic. This means that, by Lemma \ref{lemma:combined_lemma}, Point \ref{lemma:wc_intersection_makes_all_wc}, the union of any two children of $C$ is $C$ itself. Thus, any two children can be chosen to make a canonical DR-plan and that is the minimum fan-in possible for a node of the DR-plan.
% are potential choices for the optimal DR-plan as they all create equal fan-in (exactly two) at this level.

\newcommand{\induceonc}[1]{Idc\left(C,#1\right)}
\renewcommand{\induceonc}[1]{#1}
\newcommand{\iunion}[1]{\induceonc{I\cup\bigcup_{k\in [N]\setminus\{#1\}}{R_k}}}

However, to guarantee that any two are the \vemph{optimal} choice, it must ensure minimum fan-in over all descendants leading up to a common maximal antichain $S$ of subgraphs.

To prove this holds, take the set $[N]=\{1,\dots,N\}$, and denote $I=\bigcap_{k\in [N]}{C_k}$ and $R_k=C\setminus C_k$. Suppose  $C_i$ and $C_j$, where $i\neq j$,  are  the children. For convenience, we will assume all subgraphs are induced subgraphs of $C$. We know that $C=\induceonc{I\cup\bigcup_{k\in [N]}{R_k}}$ and $C_i=\iunion{i}$. The isostatic vertex-maximal subgraphs of $C_i$ are $(\iunion{i,1}),\ldots,(\iunion{i,i-1}),(\iunion{i,i+1}),\ldots,(\iunion{i,N})$ all of whose pairwise intersections are isostatic subgraphs. So any two of these are viable children for $C_i$.
% Since
% \[C_i=Idc\left(C,I\cup\bigcup_{k\in S_N\setminus\{i\}}{R_k}\right)\]
% the children of $C_i$ will be
% \[Idc\left(C,I\cup\bigcup_{k\in S_N\setminus\{i,m\}}{R_k}\right)\]
% and
% \[Idc\left(C,I\cup\bigcup_{k\in S_N\setminus\{i,n\}}{R_k}\right)\]
% % $C_i=Idc\left(C,I\cup\bigcup_{k\in S_N\setminus\{i\}}{R_k}\right)$
% % the children of this node will be
% % $Idc\left(C,I\cup\bigcup_{k\in S_N\setminus\{i,m\}}{R_k}\right)$
% % and
% % $Idc\left(C,I\cup\bigcup_{k\in S_N\setminus\{i,n\}}{R_k}\right)$
% for arbitrary $m$ and $n$, where $i,j,m,n$ do not equal each other. \todo{Prove these are valid children? Or is this obvious?}
This continues for $N-1$ levels, always with fan-in of two (the minimum possible), at which point every descendant of $C$ is some $\induceonc{I\cup R_k}$ for $k\in [N]$, with every $k$ appearing at least once. At the last level, there are exactly two rigid proper vertex-maximal subgraphs, and hence a unique choice of pair of children. Thus, regardless of the sequence of choices of $C_i$ and $C_j$, and of their descendants at each level, the DR-plan has the optimal fan-in of two for every node for  $N$ levels,  and the collection of last level nodes contain the same maximal antichain of subgraphs (for all choices).
%
% \medskip\noindent
% This proof then applies to itself recursively to show that the fan-in of the children will also be minimum.
%
% -- say this at top `proof is by induction on the level of the dr=plan'
\end{proof}

This proof of this theorem relies on the following crucial observation and lemma. These will be used again in the application sections (\ref{sec:bodypin} and \ref{sec:pinnedline}) of the paper, with modifications to work for other types of qusecs.

\begin{observation*}\label{lemma:union_intersection}
If $F_i$ and $F_j$ are non-empty isostatic graphs then the following hold: \\
(1) $F_i\cup F_j$ is not trivial;
(2) $F_i\cup F_j$ is underconstrained if and only if $F_i\cap F_j$ is trivial;
(3) $F_i\cup F_j$ is isostatic if and only if $F_i\cap F_j$ is isostatic; and
(4) $F_i\cap F_j$ is not underconstrained.
\end{observation*}

The following key properties hold at the nodes of a canonical DR-plan.

\begin{lemma*}\label{lemma:combined_lemma}
Let $C$ be an isostatic node of a canonical DR-plan, with distinct children $C_1,C_2,\ldots, C_m$. Assume $i\ne j$.
Then
\begin{enumerate}
    \item\label{lemma:wc_intersection_is_C}
    $C_i\cup C_j$ is isostatic if and only if $C_i\cup C_j = C$.

    \item\label{lemma:wc_intersection_makes_all_wc}
    If $C_i\cup C_j$ is isostatic, then $\forall k: C_i\cup C_k$ is isostatic. Alternatively, if $C_i\cup C_j=C$, then $\forall k: C_i\cup C_k=C$.

    \item\label{lemma:uc_intersection_makes_all_uc}
    If $C_i\cap C_j$ is trivial, then $\forall k: C_i\cap C_k$ is trivial.
\end{enumerate}
\end{lemma*}

\begin{remark}
The first item in the above lemma generalizes to the union of any number of children, $C_1,\ldots,C_k$, resulting in the desirable property of \dfn{cluster minimality} (defined in \cite{hoffman2001decompositionI} and in Section \ref{sec:prev}) holding for canonical-optimal DR-plans.
\end{remark}

\ClearMyMinHeight
\SetMyMinHeight{.3}{revised_c2c3_of_k33s}
\SetMyMinHeight{.7}{revised_c2c3_of_k33s_candrp}

\begin{figure*}\centering%
    \begin{subfigure}{0.3\linewidth}\centering
        \includegraphics[height=\myMinHeight]{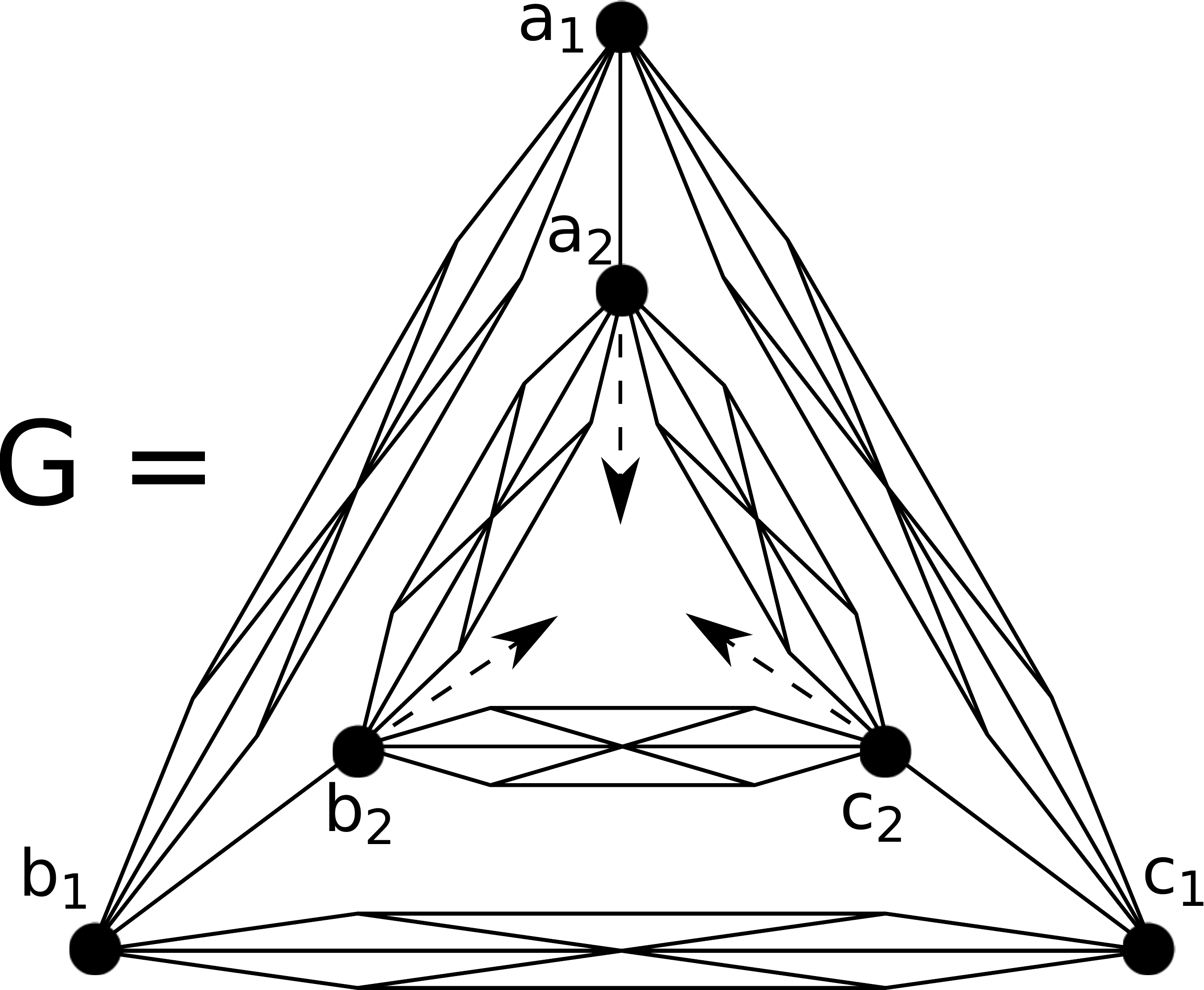}
        \caption{}\label{fig:c2c3ofk33s:a}
    \end{subfigure}%
    \hfill
    \begin{subfigure}{0.7\linewidth}\centering
        \includegraphics[height=\myMinHeight]{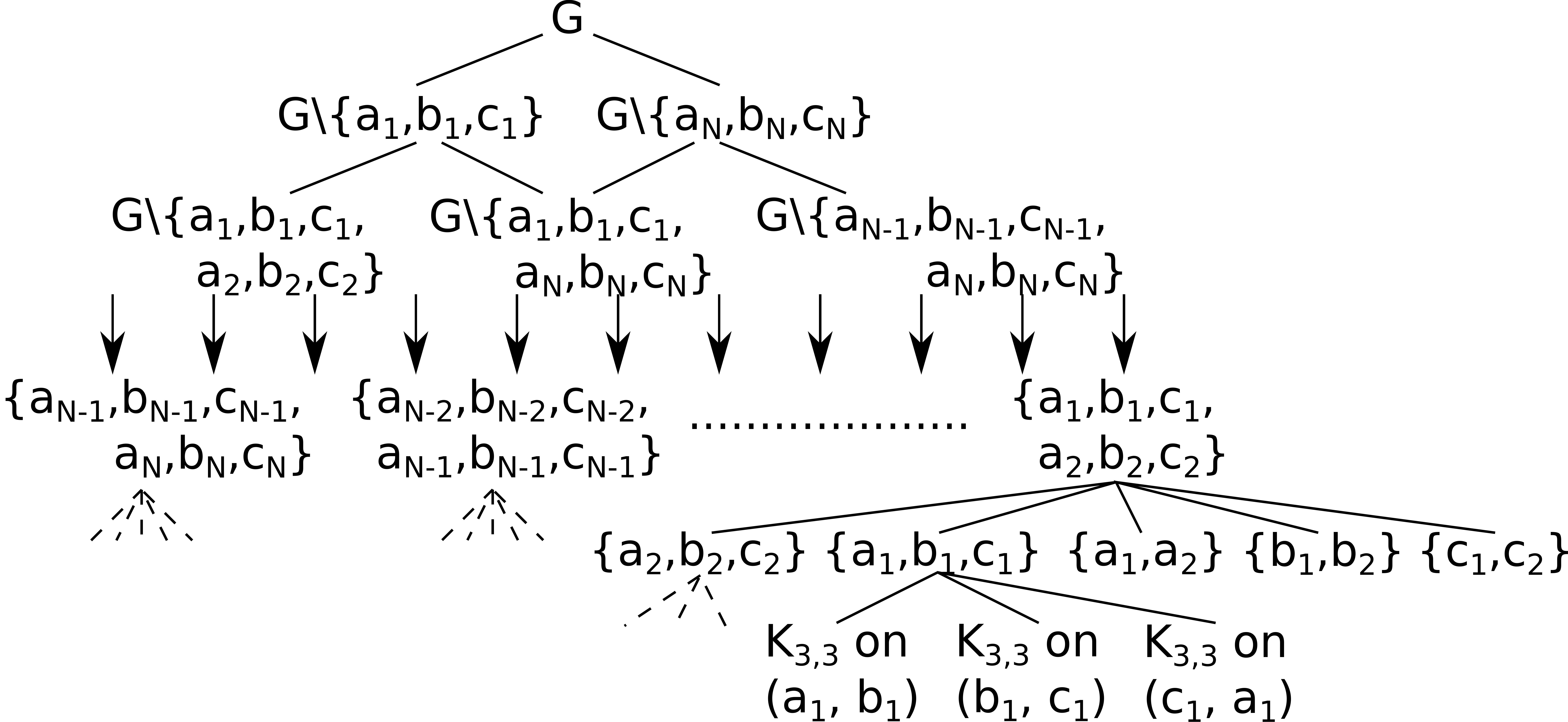}
        \caption{}\label{fig:c2c3ofk33s:b}
    \end{subfigure}%
    \caption{
    (\ref{fig:c2c3ofk33s:a}) A sequence of doublets ($C_2 \times C_3$) intersecting on triangles, where the edges of the triangles are replaced by $K_{3,3}$'s. This pattern continues inwards for a total of $N$ triangles, indicated by the dashed arrows. (\ref{fig:c2c3ofk33s:b}) The canonical DR-plan of $G$, drawn as a DAG. $G\setminus\{a_i,b_i,c_i\}$ is shorthand for $G$ difference those nodes and all of the nodes in the corresponding $K_{3,3}$ subgraphs. Below the third level, the obvious pattern continues until only the individual doublets are present (fourth level) with the ellipses indicating the remaining doublets between those shown. Decomposition of one of these doublets is shown. The dashed lines indicated that this exact decomposition (of the similar nodes on the level) is repeated. Further decomposition of $K_{3,3}$ subgraphs into the separate 9 edges is omitted from the figure.
    }
    \label{fig:c2c3ofk33s}
\end{figure*}%

\begin{example}[DR-plan for self-similar structure]
% \myexample
% \textsl{[DR-plan for self-similar structure]}
This example details the decomposition of the graph in Figure \ref{fig:c2c3ofk33s}, the canonical DR-plan of $G$. It begins with the whole (isostatic) graph as the root. The graph $G$ has only two isostatic vertex-maximal subgraphs: $G$ without the outermost triangle composed of $K_{3,3}$ graphs (triangle $1$) and $G$ without the inner triangle (triangle $N$). These intersect on $G$ without triangle $1$ and $N$ which is clearly isostatic. As explained in the proof of Theorem \ref{theorem:main}, since there are only 2 possible children, their intersection must be a node 2 levels below the parent. As expected, it is on the third level, as a child of both of $G$'s children.

Both of $G$'s children are similar to $G$, but containing only $N-1$ triangles. Therefore, the canonical DR-plans of these children follow the same pattern. This continues downward until the individual doublets are reached (there will be multiple occurrences of the same doublets at this level, but they can be represented as the same node in a DAG).

Further decomposition of one of these doublets is shown. The three edges between the triangles and the triangles themselves all intersect trivially pairwise. By Theorem \ref{theorem:main}, part 1, they must all be children in the DR-plan. Similarly, the triangles decompose into their three trivially intersecting $K_{3,3}$'s. Then the $K_{3,3}$ subgraphs decompose into their separate 9 edges.

The self-similar nature of this graph is evident in the canonical DR-plan. Many structures are repeated throughout the DR-plan, allowing for shared computation in both decomposition and recombination.
\end{example}

% \subsection{Extensions}
% This framework immediately pushes through for body-pin systems via a simple reduction. If there are $N$ pins on a body, it can be represented as a 2-tree with $N$ vertices, each corresponding to a pin, making sure to select edge distances such that the distance between pins is preserved. E.g.\ a body with two pins is an edge, three pins is a triangle, etc. Any bodies that share a pin now intersect on their vertex that corresponds to that pin. Now we have a bar-joint representation of the body-pin system in 2D and all proofs follow.

% With more effort, it can be shown that pinned line-incidence systems can also use this framework. This is done in section \ref{XXX}.

\subsection{Algorithm}

\begin{figure*}\centering%
  \includegraphics[width=0.3\linewidth]{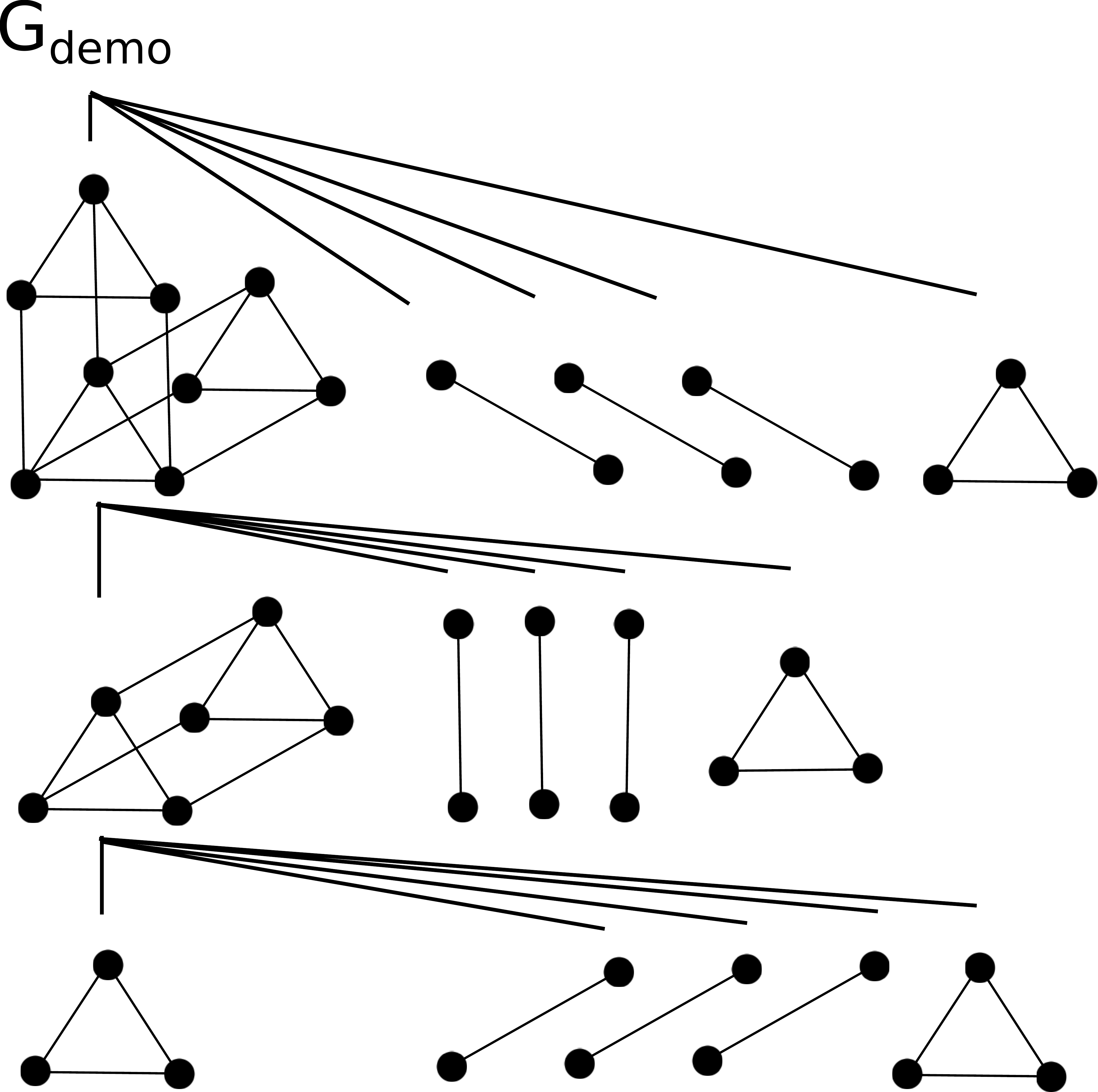}
  \caption{The sequential canonical DR-plan of $G_{demo}$ from Figure~\ref{fig:demo_graph:graph}, which is optimal (as explained in the proof of Theorem~\ref{theorem:algo_complexity}). The children of the triangle are omitted. Compare to to the typical canonical DR-plan shown in Figure~\ref{fig:demo_graph:candrp}. Also, note that the bottom-left node, the triangle, is the intersection of the 3 children of $G_{demo}$ in $ComDRP(G_{demo})$, shown in Figure~\ref{fig:demo_graph:comdrp}.}
  \label{fig:demo_graph:candrpseq}
\end{figure*}%

\begin{theorem}\label{theorem:algo_complexity}
    There exists an \candrpcomplexityv\ algorithm to find a canonical DR-plan for an input graph.
\end{theorem}

\begin{proof}
We first describe the algorithm.
The first step of the algorithm, which we call $CanDRP(G)$, is finding the isostatic vertex-maximal proper subgraphs (clusters) of the input isostatic graph $G$. One of many ways to do this is  by first dropping arbitrary edge $e$ from the edge set and running the pebble game algorithm \cite{Jacobs:1997:PG} on this subgraph, which is $O(|V|^2)$. The output of this will be a list of cluster-candidates. The next step is to isolate the set of true-clusters by running the \frontier\ algorithm \cite{hoffman2001decompositionII} \cite{lomonosov2004graph} ($O(|V|)$) on each candidate along with edge $e$, to find the minimal subgraph $D$ that contains the candidate and $e$. If $D=G$, the candidate is moved (before adding $e$) to the true-cluster list and removed from the candidate-cluster list. If $D\subsetneq G$, all candidates that are subgraphs of $D$ are removed from the candidate-cluster list and $D$ is added back to the list.  The next candidate is considered until all candidates are exhausted.

The next step is to check the intersection of any two of the true-components. If the intersection is trivial, the entire list becomes the children, and the algorithm is recursively applied to each child. If the intersection is non-trivial (necessarily isostatic), the  decomposition is computed down to $I=\bigcap_{k=1}^{N}{C_k}$. (This is discussed in detail in the proof of Theorem \ref{theorem:main}). We construct a DR-plan whose node set is that of a canonical/optimal DR-plan. Begin with the parent $C$. Its children are $S_1$ and $CanDRP(T_1)$, where $S_1=C\setminus R_1$ and $T_1$ is $R_1$, together with all incident edges. Observe that these edges are incident on vertices from $I$ and that $T_1$ is underconstrained so its DR-plan is a forest where each root becomes a child of $C$. The children of $S_1$ are $S_2$ and $CanDRP(T_2)$, where $S_2=C\setminus (R_1\cup R_2)$ and $T_2$ is $R_2$ together with all incident edges. The process iterates. At $N$ levels down, the children are  $S_N=I$ and $CanDRP(T_N)$, where $T_N$ is $R_N$ together with all incident edges. Apply the algorithm recursively to both children.
Thus we obtain an efficient method for the case of DR-plan nodes whose rigid proper vertex-maximal subgraphs have non-trivial intersections, avoiding recomputation of the isostatic components of $C$.
See Figure~\ref{fig:demo_graph:candrpseq} for an example.

To summarize the above step of the algorithm:
if the intersection is trivial, recursively apply $CanDRP$ to the entire list of true-components and set the resulting trees as the children. Else, the intersection is non-trivial, and the children of the node are $CanDRP(C_1)$ and $CanDRP(T)$, where $C_1$ is the first true-component of the parent $C$ (chosen arbitrarily, could be any child) and $T$ is the underconstrained graph formed by $C\setminus C_1$ together with all incident edges  and the associated nodes in $C_1$.

% Although the nodes containing $R_1,\ldots,R_N$ are underconstrained, simply take the children to be the roots of the forest

The DR-plan  output by this algorithm is a tree where each node is a distinct subgraph. The leaves of this tree are the edges of the starting graph $G$. Therefore, the number of nodes in the tree is $O(|E|)$ which, for isostatic input graphs, is $O(|V|)$ and the complexity of the algorithm is $O(|V|^3)$.
%
% \note In the best case, this can be $O(V^2)$ if the graph is a 2-tree
\end{proof}

\ClearMyMinHeight
\SetMyMinHeight{.6}{new_overconstrained_not_optimal}
\SetMyMinHeight{.4}{new_overconstrained_optimal}

\begin{figure*}\centering%
    \begin{subfigure}{0.5\linewidth}\centering
        \scalebox{1}[.9]{
        \includegraphics[height=\myMinHeight]{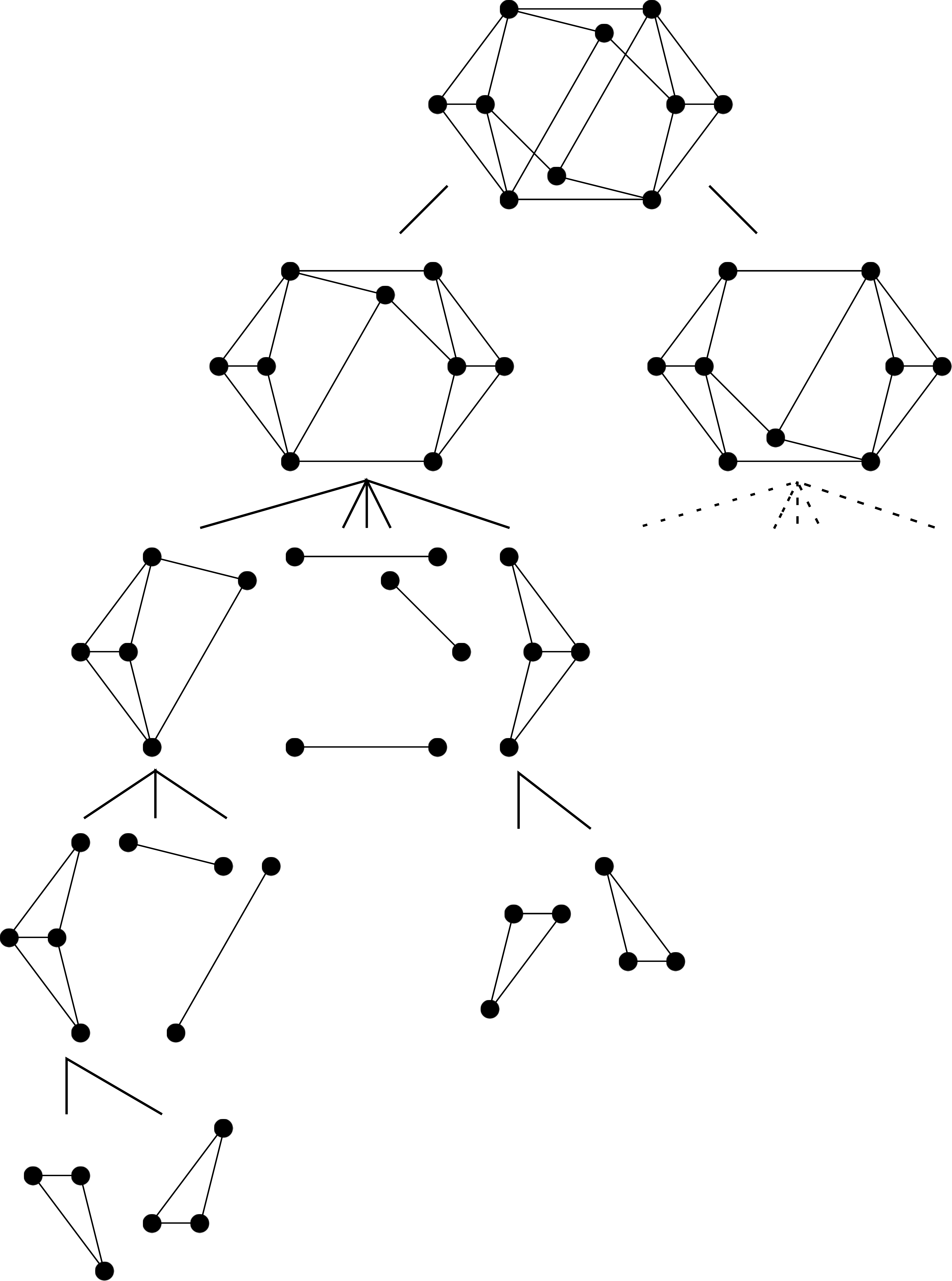}
        }
        \caption{}\label{fig:overconstrained:optimal}
    \end{subfigure}%
    \hfill
    \begin{subfigure}{0.5\linewidth}\centering
        \scalebox{1}[.9]{
        \includegraphics[height=\myMinHeight]{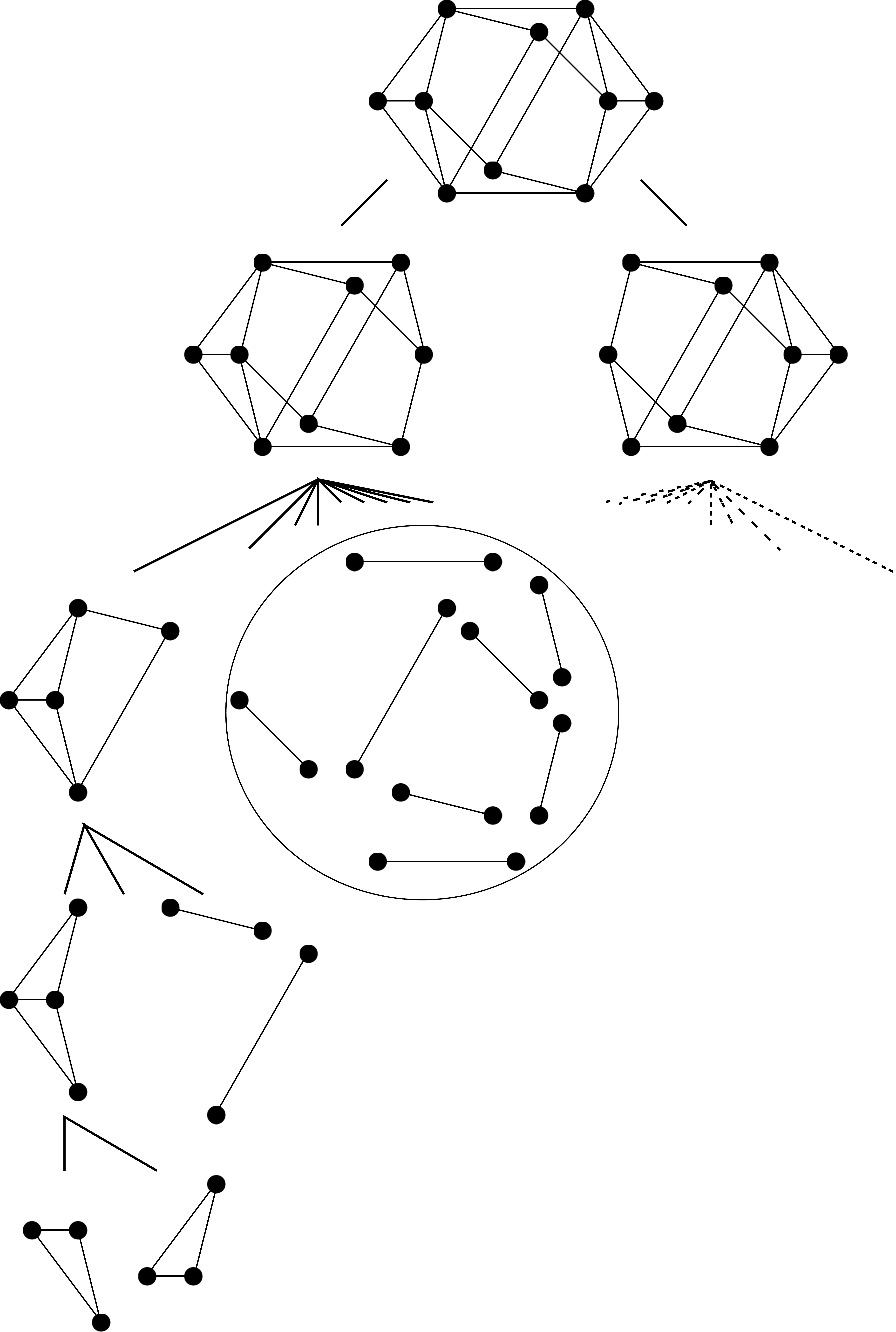}
        }
        \caption{}\label{fig:overconstrained:not_optimal}
    \end{subfigure}%
    \caption{
    % (\ref{fig:overconstrained:optimal}) An optimal DR-plan of a singly overconstrained rigid graph with a fan-in of 5. Decomposition of triangles is omitted. (\ref{fig:overconstrained:not_optimal}) A canonical and cluster-minimal DR-plan of the same graph. The DR-plan has a fan-in of 9 and is non-optimal, shown by the preceding counter-example. Since the nontrivial intersection of the two children of the root is underconstrained, their decompositions are separate, causing the large fan-in.
    Both figures are canonical and cluster-minimal DR-plans of the same singly overconstrained rigid graph. Decomposition of triangles are omitted and dashed lines indicate a decomposition similar to the other nodes on the same level. (\ref{fig:overconstrained:optimal}) is an optimal DR-plan, with a fan-in of 5. (\ref{fig:overconstrained:not_optimal}) has a fan-in of 9 and is non-optimal, shown by the preceding counter-example.
    }
    \label{fig:overconstrained}
\end{figure*}%

\subsection{Overconstrained Graphs and NP-Hardness of Optimal DR-Planning}
\label{sec:drp:overconstrained}

For overconstrained (not independent) graphs, a canonical DR-plan is still well-defined.
However, it may be far from optimal. The proofs of Theorem \ref{theorem:main}, Observation \ref{lemma:union_intersection}, and Lemma \ref{lemma:combined_lemma} all fail for overconstrained graphs.
It is important to note that, regardless whether the graph is overconstrained, if every node in a canonical DR-plan $R$ has clusters whose pairwise intersection is trivial, then the DR-plan is the unique one satisfying Property (2), and since we know that there is an optimal DR-plan that satisfies Property (2), $R$ is in fact optimal. The problem arises when some node in a DR-plan has clusters whose pairwise intersection is non-trivial.
In this case, an arbitrary choice of a pair of clusters as children of an overconstrained node in a canonical DR-plan may not result in an optimal DR-plan. This is in contrast to independent graphs, which, as shown in Theorem~\ref{theorem:main}, exhibit the strong Church-Rosser property that any choice yields an optimal DR-plan.
% \todo{rephrase? Theorem 4 shows this. If it's independent it \vemph{will} be optimal (the church rosser property), no longer holds if it's overconstrained.}
A good source of examples of overconstrained graphs with canonical DR-plans that are not optimal are graphs whose cluster-minimal DR-plans that are not optimal. The example shown in Figure \ref{fig:overconstrained} is a canonical, cluster-minimal DR-plan that is not optimal; an optimal DR-plan is also shown in the figure.
The root cause of the NP-hardness is encapsulated in this figure: because the different choices of vertex-maximal subgraphs for overconstrained input do not incur the same fan-in, finding the optimal DR-plan becomes a search problem with a combinatorial explosion of options.

As mentioned earlier, the Modified \frontier\ algorithm
version given in \cite{lomonosov2004graph} runs in polynomial time and finds a cluster-minimal DR-plan for any graph.
Similarly, the algorithm given above finds a canonical DR-plan also for any input graph.  However neither of these DR-plans may be optimal for overconstrained graphs as shown in Figure \ref{fig:overconstrained}.

While the canonical DR-plan is optimal only if the input graph is independent, when there are only $k$ overconstraints for some fixed $k$, we can still find the optimal DR-plan using a straightforward modification of the above algorithm. However, the time complexity is exponential in $k$.

This exponential growth of time complexity for overconstrained graphs is in fact captured in the proof of NP-hardness of optimal DR-planning in \cite{sitharam2005combinatorial, lomonosov2004graph}.

% \section{Using the canonical, optimal DR-plan for Realizing (Solving) Qusecs}
\section{Recombination and Problem Relationships}
\label{sec:recomb}
In this section, we consider the \dfn{optimal recombination} problem of combining specific solutions of subsystems in a DR-plan into a solution of their parent system $I$ (without loss of generality, at the top level of the DR-plan). In the case of isostatic qusecs, the parent system $I$ is isostatic (the root of the DR-plan), and we seek \vemph{solution(s) (among a finite large number of solutions) with a specific orientation or chirality}. In the case of underconstrained qusecs the subsystems are the multiple roots of the DR-plan, the parent system $I$ is underconstrained, and we typically seek an efficient algorithmic description of \vemph{connected component(s) of solutions with a specific orientation or chirality}.

The main barrier in recombination when given an optimal DR-plan  (of smallest possible size or max fan-in) for a system $S$,  is that the number of children of the root  (resp. number of roots of the DR-plan)---and correspondingly the  size and complexity of the (indecomposable) algebraic system $I$ to be solved---could be arbitrarily large as a function of the size of $S$. This difficulty can persist even after optimal parametrization of the indecomposable system $I$ as in \cite{sitharam2010optimized} to minimize its algebraic complexity.

\subsection{Previous Work}
% \todo{We now briefly survey existing techniques for studying 2D qusecs,  many of which are \dfn{bar-joint} systems (Examples 1 and 2 above, see Sections \ref{sec:prelim}, \ref{sec:DRP}, and \ref{sec:recomb}), \dfn{body-hyperpin} systems (Example 4 and 5, see Section \ref{sec:bodypin}) or \dfn{pinned-line incidence} systems (Example 3, see Section \ref{sec:pinnedline}). The limitations of these techniques directly motivate the contributions of this paper.}
We now briefly survey existing techniques for handling the complexity of recombination of DR-plans for qusecs. The limitations of these techniques directly motivate the contributions in this section.

\subsubsection{Optimal Recombination and Solution Space Navigation}
For the entire DR-plan, finding all desired solutions is barely tractable even if recombination of solved subsystems is easy for each indecomposable parent system in the DR-plan. This is because even for the simplest, highly decomposable systems with small DR-plans, the problem of finding even a single solution to the input system at the root of the DR-plan is NP-hard  \cite{saxe1979embeddability} and there is a combinatorial explosion of solutions \cite{borcea2004number}. Typically, however, the desired solution has a given orientation type, in which case, the crux of the difficulty is concentrated in the algebraic complexity of (re)combining child system solutions to give a solution to the parent system at any given level of the DR-plan. For fairly general 3D constraint systems, there are algorithms with formal guarantees that uncover underlying matroids to combinatorially obtain an optimal parameterization to minimize the algebraic complexity of the indecomposable parent (sub)systems that occur in the DR-plan \cite{sitharam2010optimized,sitharam2006well,sitharam2010reconciling}, provided the DR-plan meets some of the abovementioned criteria.

However, the generality of these algorithms trades-off against efficiency, and, despite the optimization, the best algorithms can still take exponential time in the number of child subsystems (which can be arbitrarily large even for optimal DR-plans) in order to guarantee all solutions of a given orientation type, even for a single (sub)system in a DR-plan. They are prohibitively slow in practice. We note that, utilizing the DR-plan and optimal recombination as a principled basis, high performance heuristics and software exists \cite{sitharam2006solution} to tame combinatorial explosion via user intervention.

\subsubsection{Configuration Spaces of Underconstrained Systems}
For underconstrained 2D bar-joint and body-hyperpin qusecs obtained from various subclasses of tree-decomposable systems, algorithms have been developed to complete them into isostatic systems
\cite{joan-arinyo2003transforming,sitharam2005combinatorial,gao2006ctree,sitharam2010convex} and to find paths within the connected components \cite{sitharam2011cayleyI,hidalgo2011reachability} of standard Cartesian configuration spaces. Most of the algorithms with formal guarantees leverage Cayley configuration space theory \cite{sitharam2010convex,sitharam2011cayleyI,sitharam2011cayleyII} to characterize subclasses of graphs and additional constraints that control combinatorial explosion, and provide faithful bijective representation of connected components and paths. These algorithms have decreasing efficiency as the subclass of systems gets bigger, with highest efficiency for underlying partial 2-tree graphs (alternately called tree-width 2, series-parallel, and $K_4$ minor avoiding), moderate efficiency for 1 degree-of-freedom (dof) graphs with low Cayley complexity (which include common linkages such as the Strandbeest, Limacon, and Cardioid), and decreased efficiency for general 1-dof tree-decomposable graphs. While software suites exist  \cite{keycurriculum1995geometer,porta2014open,siemens1999d,todd2007geometry}, no such formal algorithms and guarantees are known for non-tree-decomposable systems.

\subsection{Optimal Modification for Recombination}
In the following, we formulate the problem of \dfn{optimal modification} of an indecomposable algebraic system $I$ at some node of a (possibly optimal) DR-plan into a decomposable system with a small DR-plan (low algebraic complexity). Leveraging recent results on \dfn{Cayley configuration spaces}, our approach to the optimal modification problem achieves the following:

\begin{enumerate}[(a)]
    \item
    {\sl Small DR-plan.}
    We obtain a  parameterized family of systems $I_{\lambda_F}$---one for each value $\lambda_F$ for the parameters $F$,  all of which have small DR-plans. Thus, given a value $v$ for $\lambda_F$, the system $I_v$ can potentially be solved or realized easily once the orientation type of the solution is known  (when the DR-plan size is small enough).

    \item
    {\sl Solution preservation.}
    Moreover, the union of solution spaces of the systems in the family $I_{\lambda_F}$ is guaranteed to contain all of $I$'s solutions.

    \item
    {\sl Efficient search.}
    Finally,  the so-called \dfn{Cayley} or distance parameter space $\lambda_F$  is convex or otherwise easy to traverse in order to search for $I$'s solution (or connected component) of the desired orientation type. For the case when the modification (number of Cayley parameters) is bounded, this approach provides an efficient algorithm for recombination. We first define the decision version of the problem of optimal modification for decomposition. The standard optimization versions are straightforward.
\end{enumerate}

\problemstatement{Optimal Modification for Decomposition (OMD) Problem.}
Given a generically independent graph $G = (V,E)$ with no non-trivial proper isostatic subgraph (indecomposable) and 2 constants $k$ and $s$, does there exist a set of at most $k$  edges $E_1$ and a set of non-edges $F$ such that the modified graph $H = (V, E\setminus E_1 \cup F)$ has a DR-plan of size at most $s$?  The OMD$_k$ problem is  OMD where $k$ is a fixed bound (not part of the input). We say that such a tuple $(G,s)$ is a member of the set OMD$_k$. We loosely refer to graphs $G$ as OMD with appropriately small $k$ and $s$ or OMD$_k$ with appropriately small $s$.

It is immediately clear that indecomposable graphs $G$ that belong in OMD$_k$ for small $k$ and $s$  lend themselves to modification  into decomposable graphs satisfying Criteria (a) and (b) above. However, it is not clear how Criterion (c) is met by OMD graphs. Before we consider this question, we discuss previous work on recombination of DR-plans.

\subsection{Using Convex Cayley Configuration Spaces}
\label{sec:2-tree-reduction}
Next we provide the necessary background to describe a specific approach for achieving the requirements (a)--(c) mentioned above, by restricting the class of reduced graphs $G' = G\setminus E_1$ and their isostatic completions $H$ in the above definition of the OMD problem, and using a key theorem of Convex Cayley configuration spaces \cite{sitharam2010convex}. This theorem characterizes the class of graphs $H$ and non-edges $F$ (Cayley parameters), such that the set of vectors $\lambda_F$ of  attainable lengths of the non-edges $F$ is always convex for any given lengths $\delta$ for the edges of $H$ (i.e.\ over all the realizations of the bar-joint constraint system or linkage $(H,\delta)$ in 2 dimensions). This set is called the (2-dimensional) \dfn{Cayley configuration space} of the linkage $(H,\delta)$ on the Cayley parameters $F$, denoted $\Phi_F(H,\delta)$ and can be viewed as a ``projection'' of the cartesian realization space of $(H,\delta)$ on the Cayley parameters $F$. Such graphs $H$ are said to have \dfn{convexifiable Cayley configuration spaces for some parameters $F$} (short: $H$ is \dfn{convexifiable}).

To state the theorem, we first have to define the notion of \dfn{2-sums} and \dfn{2-trees}. Let $H_1$ and $H_2$ be two graphs on disjoint sets of vertices $V_1$ and $V_2$, with edge sets $E_1$ and $E_2$ containing edges $(u,v)$ and $(w,x)$ respectively.
% Let $H_1=(V_1,E_1)$ and $H_2=(V_2,E_2)$ be two graphs on disjoint sets of vertices that contain the edges $(u,v)$ and $(w,x)$ respectively.
A \dfn{2-sum} of $H_1$ and $H_2$ is a graph $H$ obtained by taking the union of $H_1$ and $H_2$ and identifying $u=w$ and $v=w$. In this case, $H_1$ and $H_2$ are called \dfn{2-sum components} of $H$. A \dfn{minimal 2-sum component} of $H$ is  one that cannot be further split into 2-sum components. A \dfn{2-tree} is recursively obtained by taking a 2-sum of 2-trees, with the base case of a 2-tree being a triangle. A \dfn{partial 2-tree} is a 2-tree minus some edges. Partial 2-trees have an alternate characterization as the graphs that avoid $K_4$ minors, and are also called series-parallel graphs.

\begin{theorem}\label{theorem:convexcayley}
    \cite{sitharam2010convex} $H$ has a convexifiable Cayley configuration space  with parameters $F$ if and only if for each $f\in F$  all the minimal 2-sum components of $H\cup F$ that contain both endpoints of $f$ are partial 2-trees. The Cayley configuration space $\Phi_F(H,\delta)$ of a bar-joint system or linkage $(H,\delta)$ is a convex polytope. When $H\cup F$ is a 2-tree, the bounding hyperplanes of this polytope are triangle inequalities relating the lengths of edges of the triangles in $H\cup F$.
\end{theorem}

The idea of our approach to achieve the criteria (a)--(c) begins with the following simple but useful theorem.

\begin{theorem}\label{theorem:omdk}
    Given an indecomposable graph $G$, let $G'$ be a spanning partial 2-tree subgraph in $G$ with $k$ fewer edges than $G$. Then $(G,2)$ belongs in the set OMD$_k$.
\end{theorem}

\begin{proof}
    The proof follows from the fact that 2-trees are well decomposable and have simple DR-plans of size 2. We know that $G$ can be reduced by removing $k$ edges to create a partial 2-tree $G'$ which can then be completed to an (isostatic) 2-tree by adding some set of non-edges $F$. Thus the modified graph $H = G'\cup F$ has  a DR-plan of size 2, proving the theorem.
\end{proof}

% \begin{proof}
% The proof follows from the fact that 2-trees are well decomposable and
% have simple DR-plans of size 2. We know that $G$ can be reduced by
% removing $k$ edges to create a partial-2-tree $G'$ which can then be
% completed to an (isostatic) 2-tree by adding some set of non-edges
% $F$. Thus the modified graph $H = G'\cup F$ has  a DR-plan of size 2,
% proving the theorem.
% \end{proof}

We refer to such graphs $G$ in short as \dfn{$k$-approximately convexifiable}, where the reduced graphs $G'$ and isostatic completions $H$ are convexifiable. As observed earlier, since graphs such as $G$ are in OMD$_k$, Criteria (a) and (b) are automatically met for small enough $k$. Criterion (c) is addressed as described in the following efficient search procedure which clarifies the dependence of the complexity on the number and ranges  of Cayley parameters $F$.

\begin{theorem*}[Efficient search]
\label{theorem:criterionc}
    % [Efficient search]
    For an indecomposable, $k$-approximately convexifiable graph $G = (V,E)$, let $G' = (V,E' =E\setminus D)$ be a spanning partial 2-tree subgraph where $|D| \le  k$. Let  $F$ be a set of non-edges of $G$ such that $H = (V, E'\cup F)$ is a 2-tree. Each solution $p$ (or connected component of a solution space) of $(G,\delta)$ of an orientation type $\sigma_p$ can be found in time $O(\log(W))$ where $W$ is the number of cells of desired accuracy (discrete volume) of the convex polytope $\Phi_F(G',\delta_E')$. The (discrete) volume $W$ is exponential in $|F|$ and polynomial in the (discrete range) of the parameters in $F$.
\end{theorem*}

\ClearMyMinHeight
\SetMyMinHeight{.2}{omd_k33_example_new-0}
\SetMyMinHeight{.15}{omd_k33_example_new-1}
\SetMyMinHeight{.15}{omd_k33_example_new-2}
\SetMyMinHeight{.15}{omd_k33_example-1}
\SetMyMinHeight{.15}{omd_k33_example-2}

\begin{figure*}\centering
    \begin{subfigure}{.2\linewidth}\centering
        \includegraphics[height=\myMinHeight]{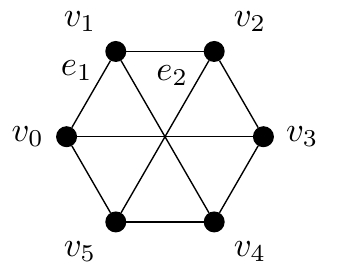}
        \caption{}\label{fig:omd_k33_example:a}
    \end{subfigure}\hfill
    \begin{subfigure}{.15\linewidth}\centering
        \includegraphics[height=\myMinHeight]{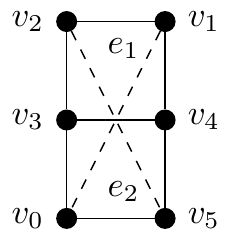}
        \caption{}\label{fig:omd_k33_example:b}
    \end{subfigure}
    \begin{subfigure}{.15\linewidth}\centering
        \includegraphics[height=\myMinHeight]{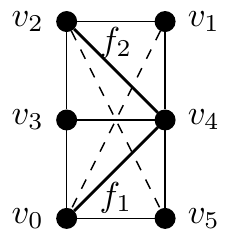}
        \caption{}\label{fig:omd_k33_example:c}
    \end{subfigure}\hfill
    \begin{subfigure}{.15\linewidth}\centering
        \includegraphics[height=\myMinHeight]{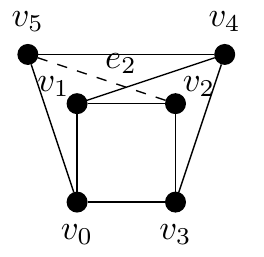}
        \caption{}\label{fig:omd_k33_example:d}
    \end{subfigure}
    \begin{subfigure}{.15\linewidth}\centering
        \includegraphics[height=\myMinHeight]{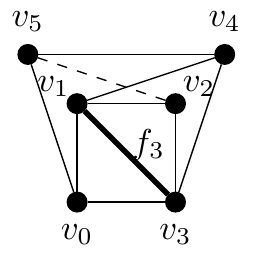}
        \caption{}\label{fig:omd_k33_example:e}
    \end{subfigure}

    \caption{
    (\ref{fig:omd_k33_example:a}) The $K_{3,3}$ with two labeled edges, $e_1$ and $e_2$. (\ref{fig:omd_k33_example:b}) The $K_{3,3}$ with $e_1$ and $e_2$ removed (dashed lines) and rearranged to illustrate that it is now a partial 2-tree. (\ref{fig:omd_k33_example:c}) The $K_{3,3}$ with $\{e_1,e_2\}$ removed and $\{f_1,f_2\}$ (bold lines) added to make a 2-tree, showing that the $K_{3,3}$ is at least OMD$_2$. (\ref{fig:omd_k33_example:d}) The $K_{3,3}$ with only $e_2$ removed (dashed line). (\ref{fig:omd_k33_example:e}) The $K_{3,3}$ with $e_2$ removed and $f_3$ (bold line) added to make a low Cayley complexity graph, showing that the $K_{3,3}$ is OMD$_1$.
    }
    \label{fig:omd_k33_example}
\end{figure*}

% \noindent
% \note A major advantage of the convex Cayley method is that it is completely unaffected when $\delta$ are intervals rather than exact values~~\cite{sitharam2010convex}.

\sidenote{A major advantage of the convex Cayley method is that it is completely unaffected when $\delta$ are intervals of values rather than exact values~~\cite{sitharam2010convex}.}
%cite my paper with Heping!!!

\begin{example}[Using Cayley configuration space]
% \myexample
% [Using Cayley configuration space].
A graph $G=K_{3,3}$  cannot be decomposed into any nontrivial isostatic graphs, i.e.\ its DR-plan has a root and 9 children corresponding to the 9 edges. Solving or recombining the system $(G,\delta)$ corresponding to the root of this DR-plan involves solving a quadratic system with 8 equations and variables. Instead of simultaneously solving this system, we could instead use the fact that $G=K_{3,3}$ is in OMD$_2$: remove the edges $e_1,e_2$ in Figure \ref{fig:omd_k33_example} to give a partial 2-tree $G'$. Now add the non-edges $f_1,f_2$ to give a 2-tree $H$ with a DR-plan of size 2. The Cayley configuration space $\Phi_f(G', \delta_{E\setminus e})$ is a single interval of attainable length values $\lambda_F$ for the edge $f$. When $\delta$ is generic, i.e.\ does not admit collinearities or coincidences in the realizations of $(G,\delta)$, the realization space of $(H, \left<\delta_{E\setminus e}, \lambda_f\right>)$ has 16 solutions $q^p_{\lambda_f}$ (modulo orientation preserving isometries), with distinct orientation types $\sigma_p$ (two orientation choices for each of the 4 triangles) that can be obtained by solving a sequence of 4 single quadratics in 1 variable (DR-plan of size 2). By subdivided binary search in the interval $\lambda_f \in \Phi_f(G', \delta_{E\setminus e})$, the desired solution $p$  of $(G,\delta)$ is found when the length of the nonedge $e$ in  the realization $q^p_{\lambda_f}$ is $\delta_e$.

In fact, we can show that $G=K_{3,3}$ is in OMD$_1$ by removing a single edge to reduce (as shown in Figure \ref{fig:omd_k33_example}) to a tree-decomposable graph of low Cayley complexity (which includes the class of partial 2-trees). In the next Section \ref{sec:tdecomp}, we discuss this issue of why the largest class of reduced graphs is desirable.
\end{example}

\subsection{Optimized Modification by Enlarging the Class of Reduced
Graphs}
\label{sec:tdecomp}
It is possible in principle to decrease $k$ for a OMD$_k$ graph (i.e.\ the number of edges to be removed to ensure an isostatic completion that is decomposable with a small DR-plan) by considering reduced graphs $G'$ (and modified graphs $H$) that come from a larger class than partial 2-trees but nevertheless have convex Cayley configuration spaces at least when the realization space is restricted to a sufficiently comprehensive orientation type. In particular, the so-called \dfn{tree-decomposable graphs of low Cayley complexity} \cite{sitharam2011cayleyI,sitharam2011cayleyII} include the partial 2-trees and many others that are not partial 2-trees. See an illustration in Figure \ref{fig:omd_k33_example}. These too result in DR-plans of size 2 or 3, putting $G$ in the class OMD$_k$ and thus meeting Criteria (a) and (b). The Criterion (c) is met---for example when $k=1$---because 1-dof Cayley configuration spaces of linkages based on such graphs $G'$ are known to be single intervals when a comprehensive orientation type $\sigma_p$ of the sought solution $p$ is given. In addition, the bounds of these intervals are of low algebraic complexity. More precisely, the bounds  can themselves be computed using a DR-plan of size 2 or 3, i.e.\ the computation of these bounds is tree-decomposable. Alternatively, the bounds are in a simple quadratic or radically solvable extension field over the rationals, or they can be computed by solving a triangularized system of quadratics.

\subsection{Problem Relationships}
\label{sec:table}

In this section we provide a unified view of the various problems studied in the previous 2 sections, along with formal reductions between them. We discuss their relationship to other known problems and results as well as open questions.

\subsubsection{Special Classes of Small DR-Plans}
As seen in the previous section, 2-trees and tree-decomposable graphs have not only small, but also special DR-plans that permit easy solving---essentially by solving a single quadratic at a time.

The \dfn{restricted optimal DR-planning problem} requires DR-plans of one of these types, which reduces to recognizing if the input graph is a 2-tree or a tree-decomposable graph for which simple near-linear time algorithms are available \cite{valdes1979recognition,fudos1997graph} and the DR-plan is a by-product output of the recognition algorithm.

In the recombination setting, the corresponding \dfn{restricted OMD$_k$  problem} requires the reduced graph $G'$ and its isostatic completion $H$ to be 2-trees as in Section \ref{sec:2-tree-reduction} or to be a low Cayley complexity tree-decomposable graph as in Section \ref{sec:tdecomp}. Clearly these problems have deterministic polynomial time algorithms in $n$, but the algorithms run in time exponential in $k$.

We discuss the complexity of the restricted OMD problem (when $k$ is part of the input) in the open-problem Section \ref{sec:futurework}.

%In the recombination setting, the corresponding \dfn{restricted
%OMD$_k$  problem} requires the reduced graph $G'$ and its isostatic
%completion $H$ to be 2-trees as in Section \ref{sec:2-tree-reduction}
%or to be a low Cayley complexity tree-decomposable graph as in Section
%\ref{sec:tdecomp}. Clearly these problems have deterministic
%polynomial time algorithms, the algorithms run in time exponential in
%$k$. As a result, when $k$ is part of the input, the \dfn{restricted
%OMD problem} has the potential to be difficult. For example, when the
%isostatic completion $H$ is required to be a 2-tree the restricted OMD
%problem is reducible to the maximum spanning series-parallel subgraph
%problem shown by \cite{cai1993spanning} to be NP-complete even if the input
%graph is planar of maximum degree at most 6. However, since the OMD
%problem has other input restrictions such as not having any proper
%isostatic subgraphs, it is not clear if the reverse reduction exists
%and hence it is unclear whether the OMD problem is NP-complete. The
%same holds for the restricted OMD problem where the isostatic
%completion $H$ is required to be a tree-decomposable graph of low
%Cayley complexity (i.e, have special, small DR-plans). One potential
%obstacle to an indecomposable graph $G$'s membership in the restricted
%OMD$_k$ for small $k$ is if $G$ is tri-connected and has large girth.
%In fact, 6-connected (hence rigid) graphs with arbitrarily large girth
%have been constructed in \cite{servatius2000rigidity}.
%
\subsubsection{Optimal Modification, Completion and Recombination: Previous Work and Formal Connections}
The OMD problem is closely related to a well-studied problem of completion of an underconstrained system to an isostatic one with a small DR-plan.
\begin{observation}\label{obs:OC_to_OMD}
    The (decision version of) the \textbf{optimal completion problem (OC)} from \cite{sitharam2005combinatorial,joan-arinyo2003transforming,zhang-gao2006well} is OMD$_0$.
\end{observation}
In fact, a \dfn{restricted OC} problem was studied by \cite{joan-arinyo2003transforming} requiring the completion to be tree-decomposable.
%It is not clear if
%the reverse reduction holds, i.e, is OMD reducible to OC?

We now connect the OMD problem to the informal \dfn{optimal recombination (OR)} problem mentioned as motivation at the beginning of Section \ref{sec:recomb}.

In order to connect the OR problem to OMD, when the input graph is the isostatic graph at the DR-plan root, we do not consider the case where the two child \dfn{solved subgraphs} (corresponding to already solved subsystems) have a nontrivial intersection (in this case the recombination is trivial). We only consider the case where no two child solved subgraphs (resp. two root subgraphs when the input graph is underconstrained) share more than 1 vertex. We replace such solved subgraphs  by isostatic graphs as follows. If a solved subgraph shares at most one vertex with the remainder of the graph, simply replace it by an edge one of whose endpoints  is the shared vertex. Otherwise, replace it by  a 2-tree graph of the shared vertices. Finally, we add the additional restriction to the OM problem that when any edge in a solved subgraph is chosen among the $k$ edges to be removed, in fact the entire solved subgraph must be removed  and all of its edges must be counted in $k$.

This reduction is used also for adapting  algorithms for optimal DR-planning, recombination, completion, OMD, and other problems from bar-joint systems to so-called \dfn{body-hyperpin}, defined in Section \ref{sec:bodypin}, by showing that the problems for the latter are reduced to the corresponding problems on bar-joint systems.

%\note This section deals only with bar-joint systems, however,
%extensions of several of these problems, approaches and solutions to
%other qusecs beyond  bar-joint - specifically pinned line incidence
%systems and multi-body-pin systems will be expanded in Sections
%\ref{sec:pinnedline} and \ref{sec:bodypin}.

\section{Application: Finding Completions of Underconstrained Glassy Structures from Underconstrained to Isostatic}
\label{sec:bodypin}

% \bibliography{paper}

\ClearMyMinHeight
\SetMyMinHeight{.3}{Silica}
\SetMyMinHeight{.2}{Silicon_tetrahedron}

\begin{figure}\centering
    \includegraphics[height=\myMinHeight]{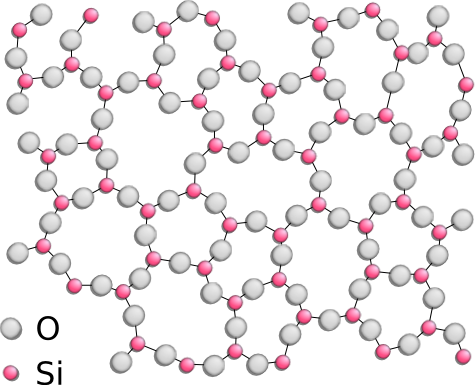}
    \hspace{0.5cm}
    \includegraphics[height=\myMinHeight]{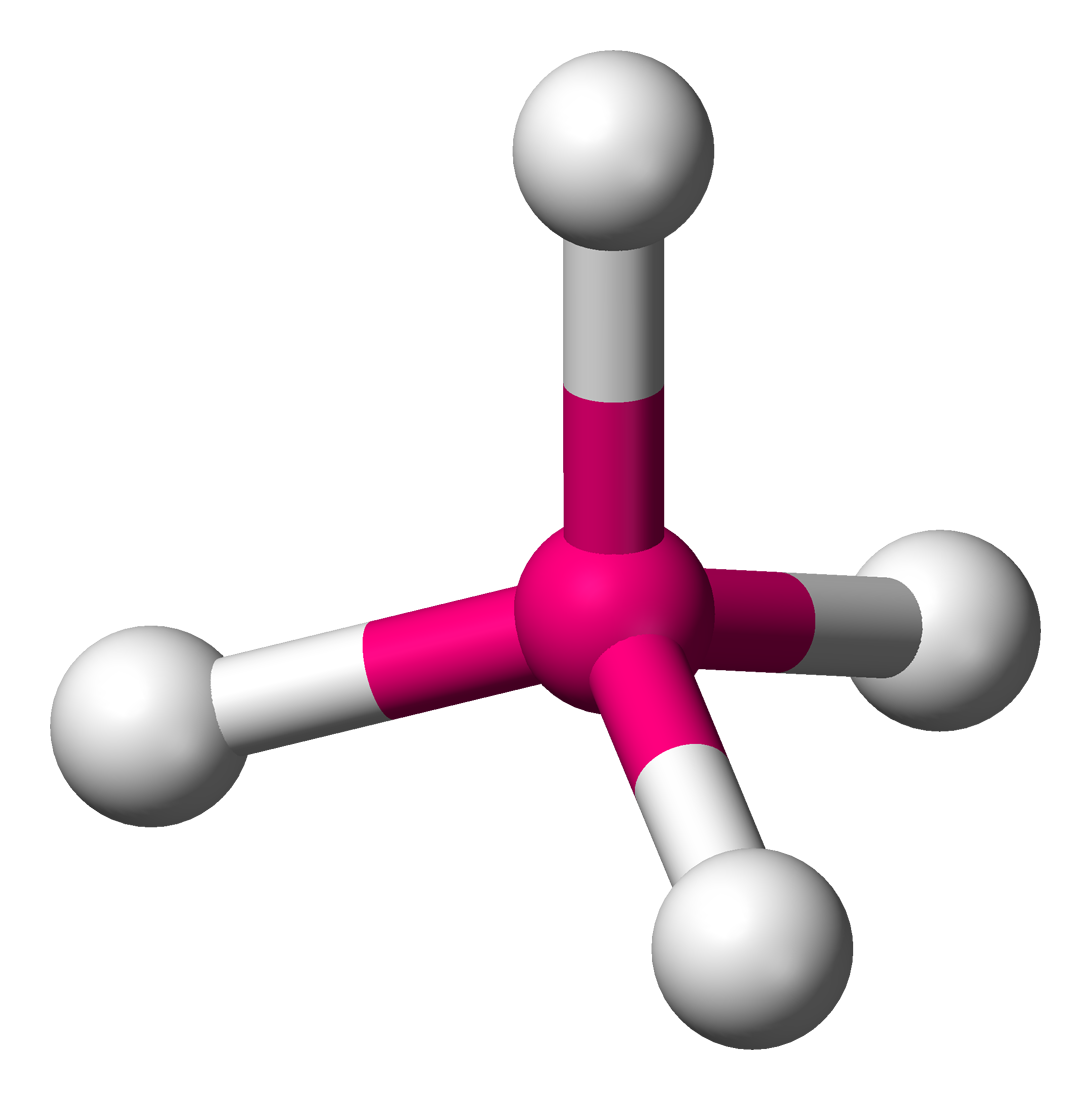}
    % \raisebox{0.25\height}{
    %     \includegraphics[height=\myMinHeight]{Silicon_tetrahedron}
    % }
    \caption{Example of a single monolayer of a silica (silicon and oxygen) glassy structure. This can be viewed as a triangle-multipin qusecs where the silicon atoms are the triangles and the oxygen atoms are the pins. Not shown here are other monolayers stacked on this one. Each silicon atom has the structure seen to the right, and so binds to another oxygen atom in an adjacent monolayer. Pictures taken from \cite{silica_figure} and \cite{tetra_silica_figure}.}
    \label{fig:silica_glass}
\end{figure}

\begin{figure}\centering
    \begin{subfigure}{0.35\linewidth}
        $
        \vcenter{\hbox{\includegraphics[width=.4\linewidth]{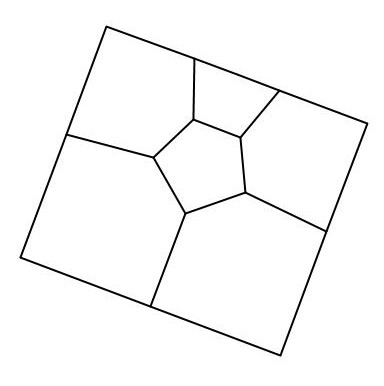}}}
        % \raisebox{-.5\height}{\scalebox{1}{$\rightarrow$}}
        \vcenter{\hbox{$\rightarrow$}}
        \vcenter{\hbox{\includegraphics[width=.4\linewidth]{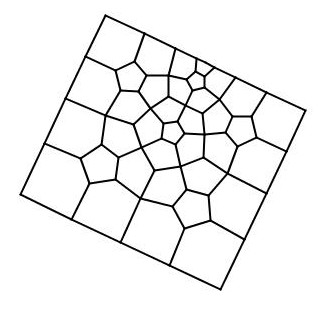}}}
        $
        \caption{}\label{fig:subdivision:simple}
    \end{subfigure}
    \begin{subfigure}{0.55\linewidth}
        \includegraphics[width=\linewidth]{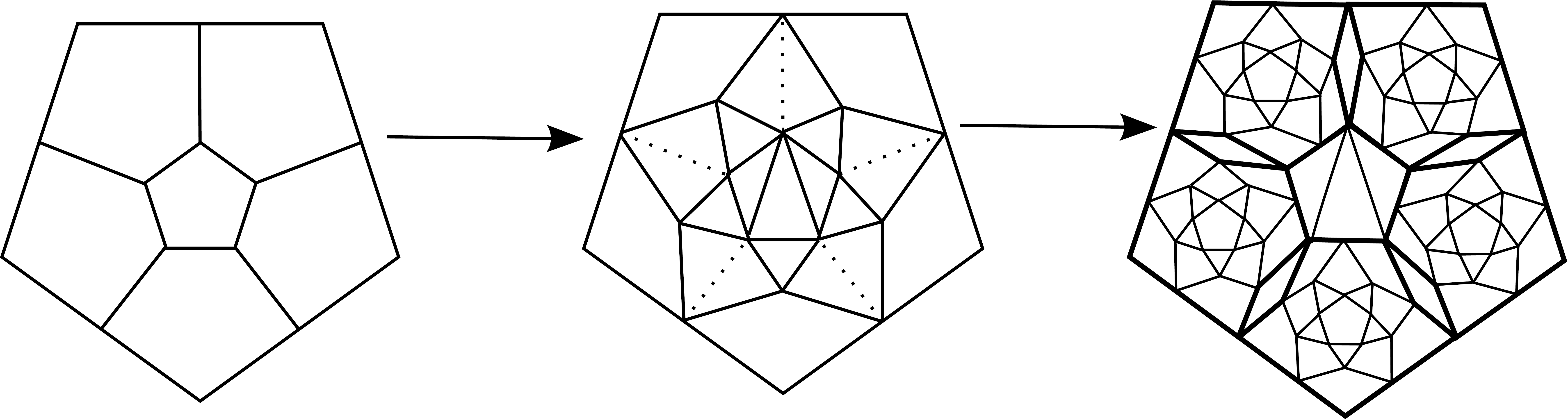}
        \caption{}\label{fig:subdivision:complicated}
    \end{subfigure}
    \caption{Examples of self-similarity via repeated subdivision. In (\ref{fig:subdivision:simple}) \cite{subdivision_paper}, there is a simple subdivision scheme that does not guarantee isostaticity. In (\ref{fig:subdivision:complicated}), a more complicated scheme is used, ensuring that the resulting graph is isostatic but not  tree-decomposable.}
    \label{fig:subdivision}
\end{figure}

We can use qusecs DR-plans to design materials such as disordered graphene and silica bi-layers \cite{silica_bilayers} \cite{sructure_of_2d_glass}. We investigate a more specific problem in a somewhat more general setting: the problem of finding boundary conditions (additional constraints) to add to an underconstrained monolayer to make it isostatic. This can be done in a number of ways: (1) pin together 2 underconstrained monolayers in such a way that the resulting bi-layer becomes isostatic (see Figure \ref{fig:silica_glass}); (2) pin the boundary of (or in general, add constraints to) a layer (possibly a genus 0 monolayer) so that it becomes isostatic; or (3) design a broader class of structures to ensure they are isostatic, self-similar (via some subdivision rule) and in addition isostatic at each level of the subdivision (see Figure \ref{fig:subdivision}).

In all cases, we are specifically interested in how to add additional constraints such that the resulting isostatic structure has a small DR-plan; this way a realization can be found, allowing efficient stress, flex and other property design related to the rigidity matrix. To answer these questions, we first introduce the qusecs' that are used to model these materials. In this section, we discuss Item (2) in detail.

% \begin{figure}\centering
%     (a)
%     \begin{subfigure}{0.2\linewidth}
%         \includegraphics[width=\linewidth]{pentl1}
%     \end{subfigure}
%     %
%     $\rightarrow$
%     \begin{subfigure}{0.2\linewidth}
%         \includegraphics[width=\linewidth]{pentl2}
%     \end{subfigure}
%     %

%     (b)
%     \begin{subfigure}{0.8\linewidth}
%         \includegraphics[width=\linewidth]{pentawesome}
%     \end{subfigure}
%     \caption{Examples of self-similarity via repeated subdivision. In (a), simple subdivision that does not guarantee isostaticity. In (b), a more complicated scheme is used, ensuring that the resulting graph is isostatic but not  tree-decomposable. Credit to \cite{subdivision_paper} for (a).}
%     \label{fig:subdivision}
% \end{figure}

\subsection{Body-Hyperpin Qusecs}

\begin{definition}
    A \dfn{body-hyperpin qusecs} is a constraint system where the objects are rigid bodies,  subsets of which are pinned together by pins; i.e, are incident at a common point.
\end{definition}

\begin{remark*}\label{rem:bodypin_is_barjoint}
    A body-hyperpin qusecs is a special case of bar-joint qusecs of the previous sections of the paper. As such, the DR-planning for isostatic systems discussed in Section \ref{sec:DRP} is unchanged and the results of Section \ref{sec:recomb} still go through with minor modifications.
\end{remark*}

For the remainder of this section, we deal only with the DR-plan of such qusecs. Hence, we refer only to the combinatorics or underlying hypergraph of the qusecs. We now introduce 2 sub-classes of body-hyperpin graphs for modeling Examples 4 and 5 in Section \ref{sec:intro}, for which the optimal completion problem is significantly easier.

\begin{figure*}\centering
    \begin{subfigure}{0.2\linewidth}\centering
        \includegraphics[height=\myMinHeight]{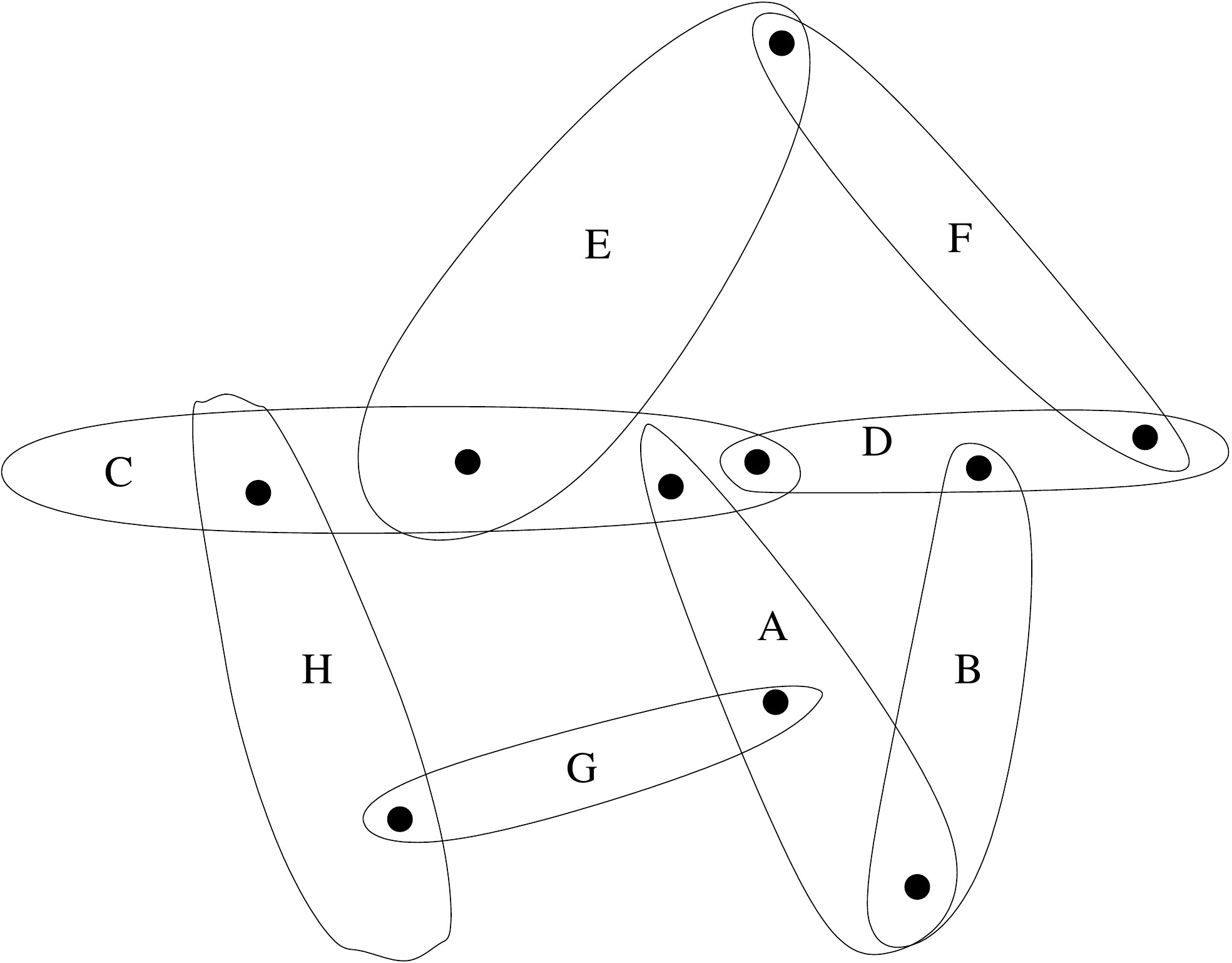}
        \caption{}\label{fig:bodypindrp:a}
    \end{subfigure}
    \hfill
    \begin{subfigure}{0.2\linewidth}\centering
        \includegraphics[height=\myMinHeight]{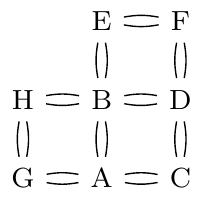}
        \caption{}\label{fig:bodypindrp:b}
    \end{subfigure}
    \hfill
    \begin{subfigure}{0.35\linewidth}\centering
        \includegraphics[height=\myMinHeight]{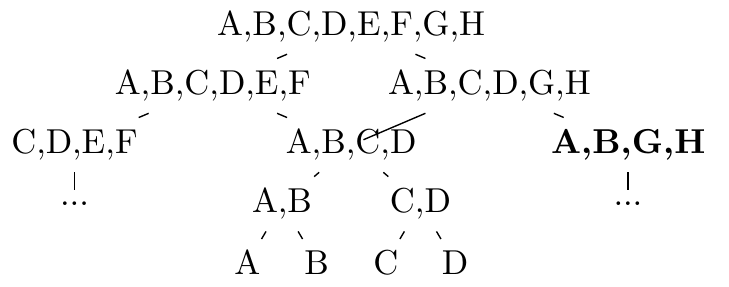}
        \caption{}\label{fig:bodypindrp:c}
    \end{subfigure}
    \hfill
    \begin{subfigure}{0.2\linewidth}\centering
        \includegraphics[height=\myMinHeight]{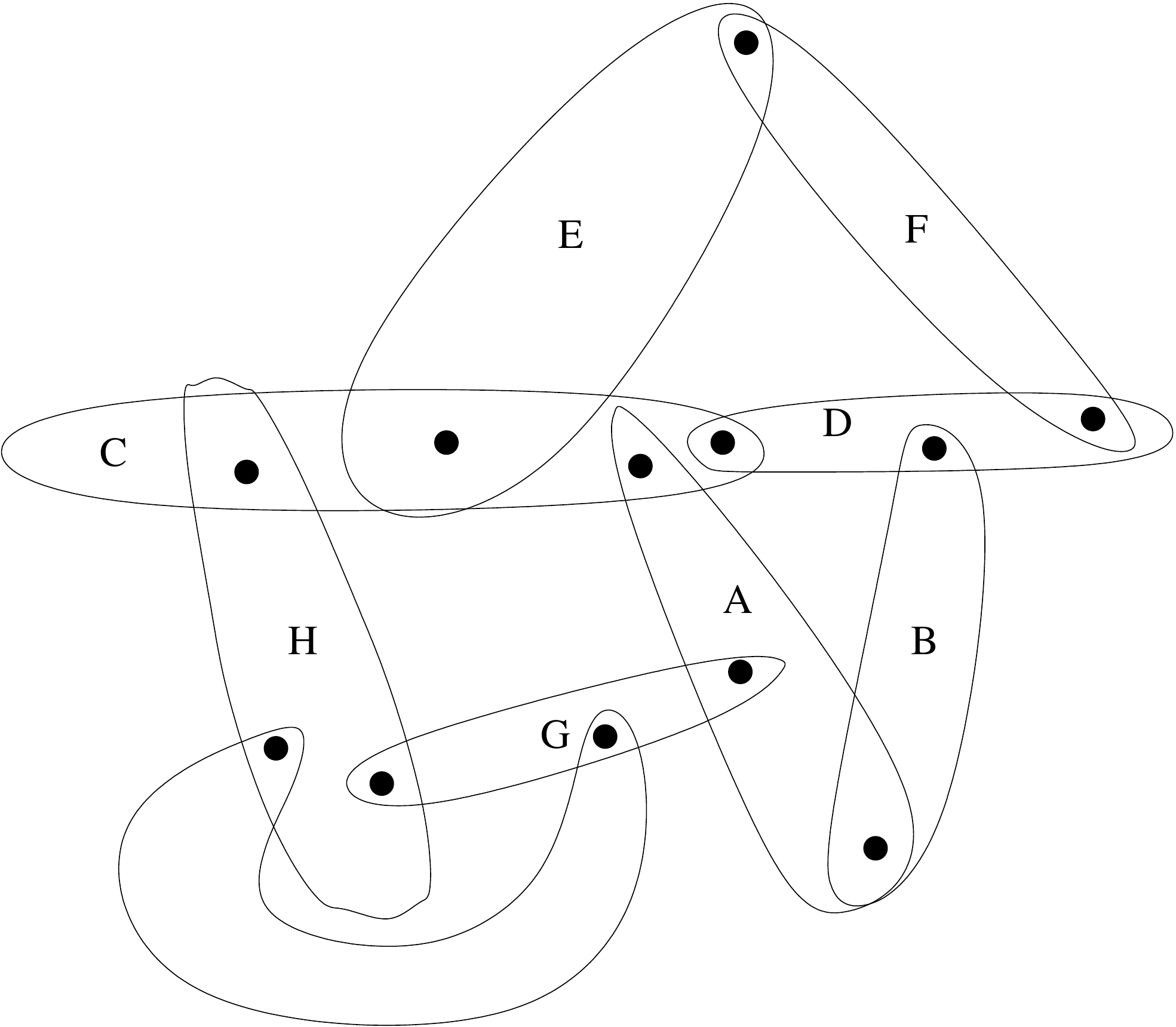}
        \caption{}\label{fig:bodypindrp:d}
    \end{subfigure}
    \caption{(\ref{fig:bodypindrp:a}) A 1-dof body-pin graph. (\ref{fig:bodypindrp:b}) The corresponding body-bar graph, explained in Definition \ref{def:body-pin}. (\ref{fig:bodypindrp:c}) The 1-dof DR-plan for the graph. In this case, to obtain an isostatic system, we would need to add a body and 2 pins to one of the nodes in the second level. (\ref{fig:bodypindrp:d}) The result of adding one such body to the bold-faced node.}
    \label{fig:bodypindrp}
\end{figure*}

\begin{definition}\label{def:body-pin}
    A body-pin graph is a body-hyperpin graph with the following conditions:
    (1) each pin is shared by at most two bodies; and
    (2) no two bodies share more than one pin

    Such a body-pin graph, $G_{BP}$, can also be seen as a \dfn{body-bar graph}, $G_{BB}$, where the bodies of $G_{BB}$ are the original bodies of $G_{BP}$ and each pin between bodies in $G_{BP}$ are replaced with 2 bars in $G_{BB}$ between the same bodies.
    Such body-bar graphs with 1 and 2-dof can be characterized by being $(3,4)$ and $(3,5)$-tight respectively \cite{Lee:2007:PGA} \cite{streinu2009sparse} (defined in \ref{sec:appendix:defs}). See Figure \ref{fig:bodypindrp:b}.
\end{definition}

\begin{definition}
    A \dfn{triangle-hyperpin graph} is a body-hyperpin graph where each body is a triangle, i.e., it shares  pins with at most 3 other bodies. This is also represented as a hyper-graph where each pin is a vertex and each triangle represents a tri-hyperedge. For such hypergraphs, {\em 1 and 2-dof} can be characterized by $(2,4)$ and $(2,5)$-tightness respectively \cite{Lee:2007:PGA} \cite{streinu2009sparse}.
\end{definition}

Body-pin graphs are of particular interest to us in the context of Example 4 in Section \ref{sec:intro}. Triangle-multipin graphs can be used to represent the silica bi-layers and glassy structures described in Example 5 of Section \ref{sec:intro}, where each triangle is the junction of ``disks'' in the plane (see Figure \ref{fig:silica_glass}). Typically, these systems are not isostatic, so to relate the work of this paper to the systems, we define a slightly different kind of DR-plan using the notion of $(k,l)$-sparsity and tightness.

\begin{definition}
    A \dfn{$(k,l)$-tight DR-plan} is one in which each child node is either a vertex maximal proper $(k,l)$-tight sub-graph of the parent node or it is trivial. In our case, the trivial nodes are the bodies.
\end{definition}

Provided such $(k,l)$-sparse graphs are matroidal (conditions given in \cite{Lee:2007:PGA}),
the notion of a canonical DR-plan extends directly to the case when the hypergraph is $(k,l)$-sparse (i.e., independent) using the straightforward notion of trivial and non-trivial intersections and $(k,l)$-tightness conditions as in Section \ref{sec:DRP}. In particular, we define canonical DR-plans with similar properties for the 1 and 2-dof body-pin and triangle-hyperpin systems defined above.

\begin{observation*}
\label{obs:bodypin_drp}
For the 1-dof body-pin graphs described above that are $(3,4)$-sparse, a $(3,4)$-tight canonical DR-plan exists where every node of a $(3,4)$-sparse graph satisfies one of the following: (1) its children are 2 proper vertex-maximal 1-dof graphs that intersect on another 1-dof graph; or (2) its children are all of the proper maximal 1-dof sub-graphs, pairwise sharing at most one body.
\end{observation*}

% \begin{observation}
% \label{obs:bodypin_drp}
%     For the 1-dof body-pin graphs described above, $(3,4)$-sparsity corresponds to a canonical DR-plan whose nodes satisfy one of the following, either
%     (a) the children are 2 proper vertex maximal 1-dof graphs that intersect on another 1-dof graph, or
%     (b) the children are a number of proper maximal 1-dof sub-graphs, sharing at most one body.
% \end{observation}

% \begin{proof}
%     \todo{Item 1 Follows from paper?}

%     For item 2, consider the case where we have more than 2 proper vertex maximal 1-dof sub-graphs $s_1, ..., s_k, k > 2$. Then, if $k_i$ and $k_j$ are joined by $2$ pins, $k_i \cup k_j$ would be $(3,4)$-tight and hence $k_i$ and $k_j$ are not vertex maximal.
% \end{proof}

As in Section \ref{sec:DRP},  a strong Church-Rosser property holds, making all canonical DR-plans optimal:
\begin{observation*}\label{rem:1dofcanon}
    When the input is independent, all $(3,4)$-tight canonical DR-plans are optimal. We can find such a DR-plan in the same time complexity as the $(2,3)$-tight case for bar-joint graphs discussed in Section \ref{sec:DRP}.
\end{observation*}

The above-mentioned algorithm exists because such $(3,4)$-tight graphs are matroidal and have a pebble game \cite{Lee:2007:PGA}.

\ClearMyMinHeight
\SetMyMinHeight{.2}{bodypin}
\SetMyMinHeight{.2}{bodypin_graph}
\SetMyMinHeight{.35}{bodypin_drp}
\SetMyMinHeight{.2}{bodypin2}

The above discussion leads to the main theorem:
\begin{theorem*}
\label{thm:1dofcase}
    Given a 1-dof body-pin or triangle-multipin graph and corresponding 1-dof DR-plan, there is a quadratic algorithm for the 1-dof optimal completion problem of Section \ref{sec:recomb}.
\end{theorem*}

\begin{observation*}
\label{obs:2dof_case}
    For the 2-dof case, provided an analogous statement to Observation \ref{rem:1dofcanon} holds, then Theorem \ref{thm:1dofcase} holds for the 2-dof  systems.
\end{observation*}

% \begin{proof} (of Observation \ref{obs:2dof_case})
%     The only difference from the 1-dof case is that now we need to remove 2-dof from our graph. We build a 2-dof DR-plan $T$. Like above, we need to add a body and 2 pins to 2 nodes now to get to isostatic.

%     Suppose we pin 2 distinct nodes $v_i$ and $v_j$. Then, there will be some common ancestor $a$ of $v_i$ and $v_j$. Then, in $T_{v_i,v_j}$, $fanin_{v_i,v_j}(a) = nl(T^a)$. However, if we chose to pin one of $v_i$ and $v_j$ twice, then $fanin_v(a) = nl(T^a) - nl(T^{a'}) +1$ . Thus $fanin_v(a)' \leq fanin_{v_i,v_j}(a)$. All ancestors of $a$ are unchanged. So $|T_v| \leq |T_{v_i,v_j}|$.

%     So, we will always end up pinning a single node twice. Hence, we can run the same algorithm as the 1-dof case and just pin twice instead of once.
% \end{proof}

\begin{remark}
\label{obs:algebraic_completion}
    While the proof for Theorem \ref{thm:1dofcase} gives us a DR-plan for the isostatic completion with minimum fan-in, (a reasonable measure of algebraic complexity), a more nuanced measure that treats solutions of 1-dof and 2-dof systems as 1 or 2 parameter families would no longer be optimized by the algorithm given in that proof. In particular, the complexity of the standard algorithm in the $k$-dof case would be exponential in $k$ (even if the case were matroidal and an optimal DR-plan is known).
\end{remark}

\section{Application: Finding Optimal DR-Plans and Realizations for Cross-Linking Microfibrils }
\label{sec:pinnedline}

The canonical DR-plan of Section \ref{sec:DRP} can additionally be applied to analyze and solve the structure of cross-linking collagen microfibrils in animals, cellulose microfibrils in plant cell walls, and actin filaments in the cytoskeleton by modeling these structures as a third type of qusecs, \dfn{pinned line-incidence systems}.

\begin{figure}\centering%
\begin{subfigure}{0.32\linewidth}\centering
    \includegraphics[height=\myMinHeight]{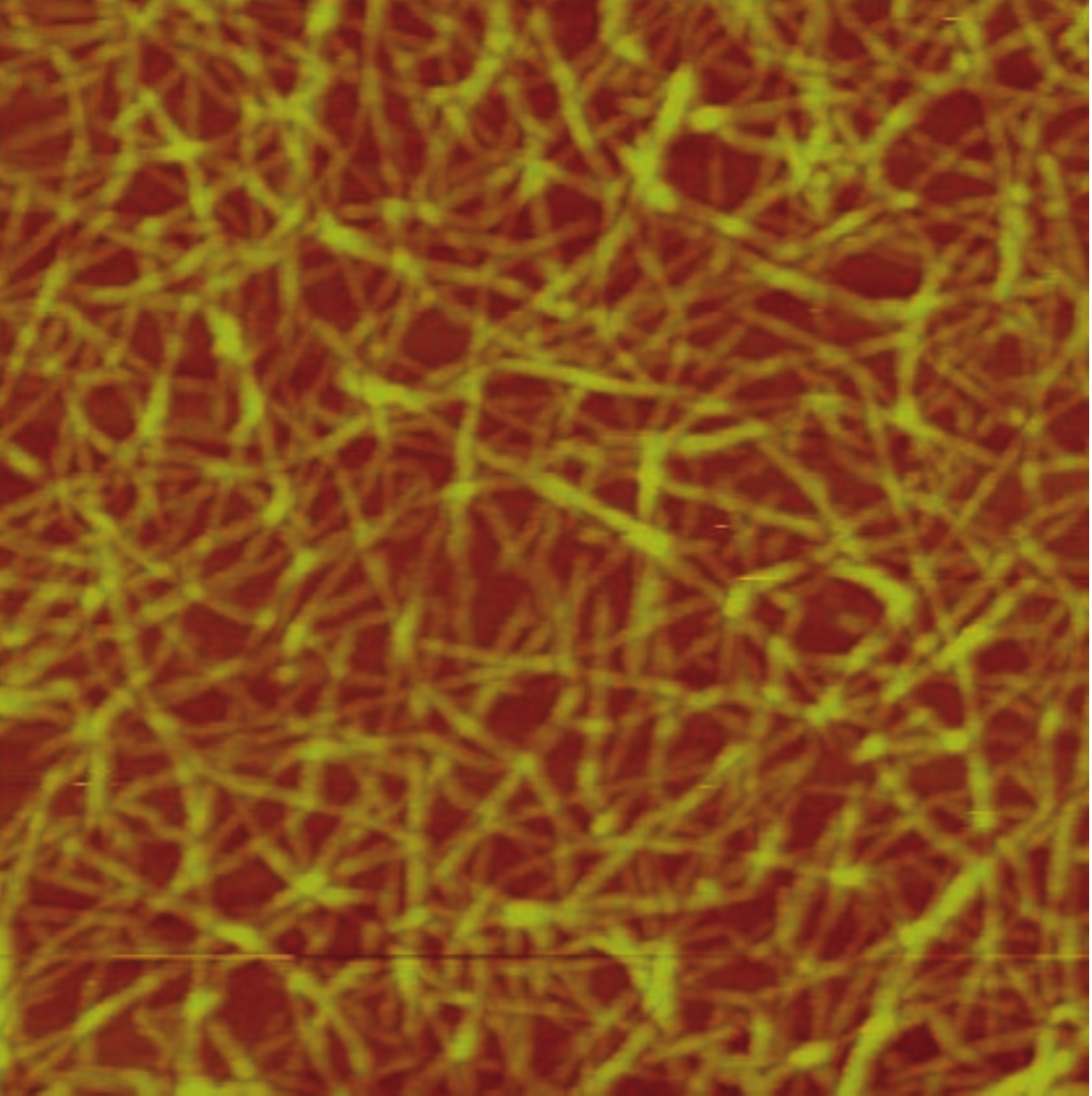}
    \caption{}\label{fig:cellulose}
\end{subfigure}\hfill%
\begin{subfigure}{0.32\linewidth}\centering
    \includegraphics[height=\myMinHeight]{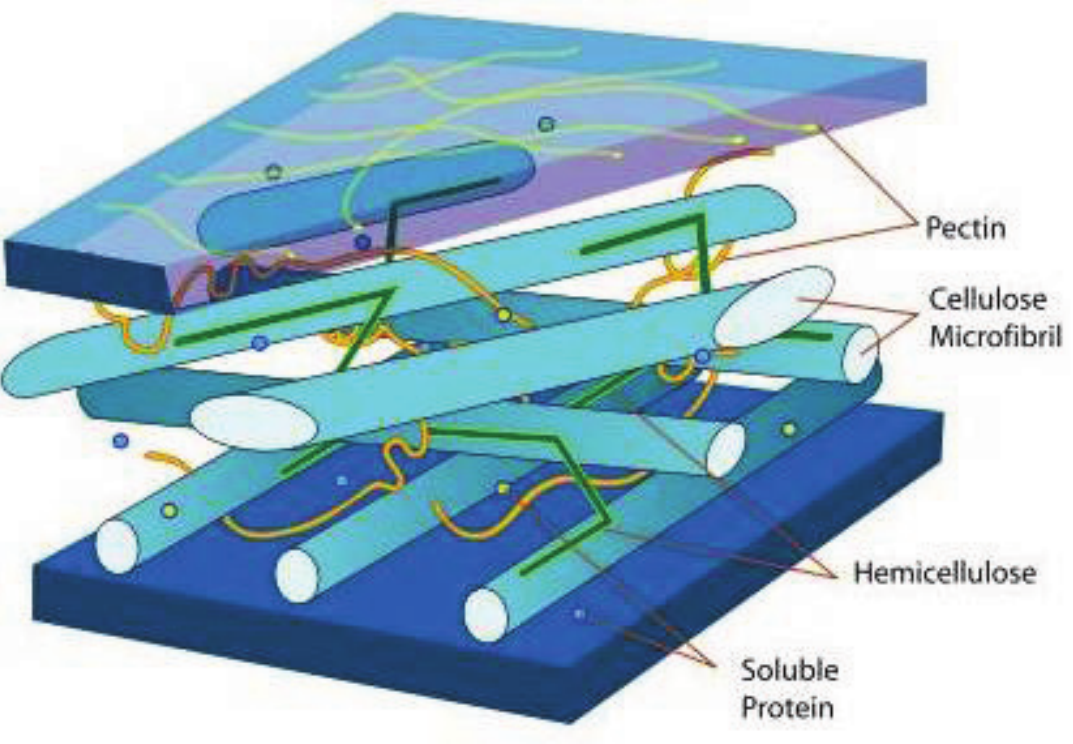}
    \caption{}\label{fig:crosslink}
\end{subfigure}\hfill%
\begin{subfigure}{0.32\linewidth}\centering
    \includegraphics[height=\myMinHeight]{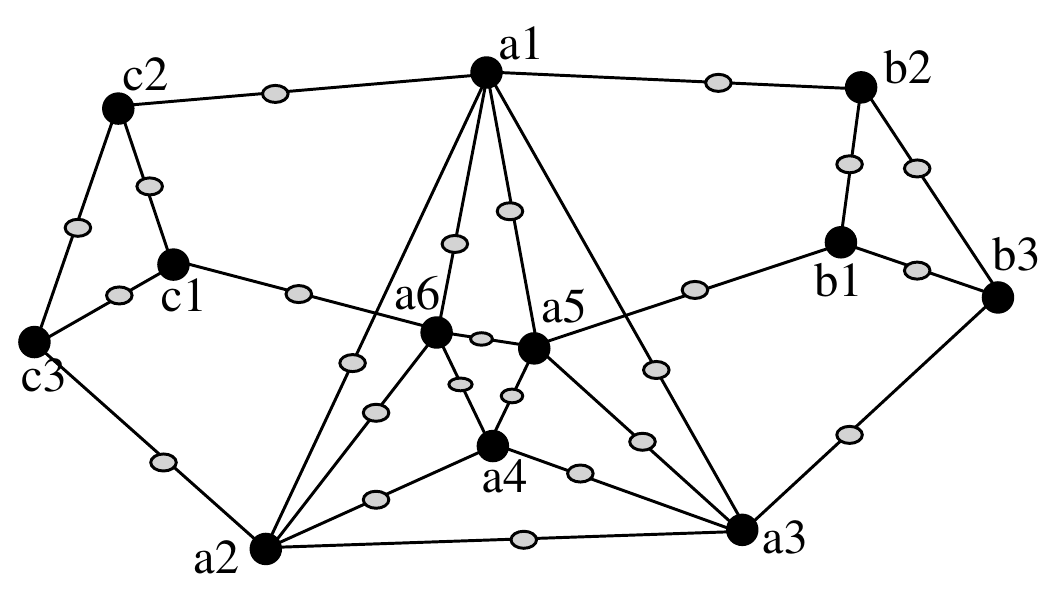}
    \caption{}
    \label{fig:pinned_line}
\end{subfigure}%
\caption{(\ref{fig:cellulose}) Microfibrils of carboxymethylated nanocellulose adsorbed on a silica surface \cite{wikimediacommons2010afm}. \ref{fig:crosslink} Cross-linking of cellulose microfibrils \cite{wikimediacommons2007plant}. \ref{fig:pinned_line} An isostatic pinned line-incidence graph, representing fibrils and their attachments.}
\end{figure}

\dfn{Collagen} is an important protein material in biological tissues with highly elastic mechanical properties \cite{buehler2008nanomechanics}. \dfn{Cellulose} is the most important constituent of the cell wall of plants (see Figure~\ref{fig:cellulose}) \cite{fall2013physical,smith1971plant}. Both of these substances consist of a large number \dfn{microfibrils},
%fibril itself straight?
%Each fibril is attached to some fixed larger organelle/membrane at one spot (typically on the boundary).
%It is additionally
each of which is cross-linked at 2 places with usually 3 other fibrils, where the \dfn{cross-linking} is like an incidence constraint that the crosslinked fibrils can slide against each other while remaining incident (see Figure~\ref{fig:crosslink}).

%Collagen
%
%layers of microfibril with cross-linking???
%
%
%- different cross-linking densities
%larger cross-link densities lead to larger yield strains, larger yield stresses as well as larger fracture stresses.
%the maximum fracture stress of the collagen fibril does not increase with increasing cross-link densities (2 links per molecule)
%
%
%Plant cell wall, cellulose
%
%Non-charged cellulose is considered to form a random mesh, the rigidity of which would depend on the number of interfibriller cross links, i.e. H bonds.
%new microfibrils were laid down on top of the existing ones, forming a new layer.

% \begin{figure}
% \centering

% \begin{subfigure}{.49\linewidth}
%   \centering
%   \includegraphics[width=.8\linewidth]{AFM_Innventia_nanocellulose}
%   \caption{}
%   \label{fig:cellulose}
% \end{subfigure}
% \begin{subfigure}{.49\linewidth}
%   \centering
%   \includegraphics[width=\linewidth]{crosslink}
%   \caption{}
%   \label{fig:crosslink}
% \end{subfigure}%
% \caption{(a) Microfibrils of carboxymethylated nanocellulose adsorbed on a silica surface \cite{wikimediacommons2010afm}. (b) Cross-linking of cellulose microfibrils \cite{wikimediacommons2007plant}.}
% \end{figure}

\ClearMyMinHeight
\SetMyMinHeight{.32}{AFM_Innventia_nanocellulose}
\SetMyMinHeight{.32}{crosslink}
\SetMyMinHeight{.32}{pinned}

\subsection{Modeling the Fibrils as a Pinned Line-Incidence System}

The cross-linking microfibrils can be modeled as a pinned line-incidence constraint system in $\mathbb{R}^2$, where incidence constraints are used instead of distance constraints.

A \dfn{pinned line-incidence system} $(G,\delta)$ is a graph $G=(V,E)$ together with parameters $\delta$ specifying $|E|$ \textbf{pins} with fixed positions in $\mathbb{R}^2$, such that each edge is constrained to lie on a line passing through the corresponding pin, i.e.\ $\delta: E \rightarrow \mathbb{R}^2$.
%
% $G=(V,E, w_V, w_E)$
%%is a \textbf{2-dimensional pinned line-incidence graph},
% has weight functions $w_V(v) = 2, w_E(e) = 1$ for all $v \in V, e \in E$.
%Additionally, there is  a  set $X$ of $|E|$ \textbf{pins} with fixed positions in $\mathbb{R}^2$,
%such that  %every edge $e \in E$ corresponds to a pin $x \in X$ where
%for each edge $e \in E$, there is a uniquely corresponding pin $x \in X$,
%such that $e$ is constrained to lie on a line passing through $x$.
%%There exists a set $X$ of $|E|$ pins fixed in $\mathbb{R}^2$, such that each edge $e_i$
%
A pinned line-incidence graph $G$ is rigid if $|E| = 2|V|$ and $|E'| \le 2|V'|$ for every induced subgraph $(V',E')$ \cite{sitharam2014incidence}. Note that no trivial motion exists since the pins have fixed positions on the plane. Euclidean transformations are not factored out. In particular, both a single vertex and a single edge are underconstrained graphs.

In the case of microfibril cross-linking, each fibril is attached to some fixed larger organelle/membrane at one site. Consequently, each fibril can be modeled as an edge of the graph, with the attachment being the corresponding pin. The two cross-linkings in which the fibril participates are modeled as the two vertices in $V$ defining the edge.

Figure~\ref{fig:pinned_line} shows an example of a pinned line-incidence graph, where the grey ovals denote pins representing attachments of fibrils, and the vertices $a_1,a_2,\ldots, c_3$ represent cross-linkings.
% and the black dots stand for vertices/cross-linkings.
%There are  and 10 vertices/cross-linkings in the graph.
The graph is isostatic, with 12 vertices and 24 edges/pins.

% \begin{figure}\centering
%     \includegraphics[width=.7\linewidth]{pinned}
%     \caption{An isostatic pinned line-incidence graph.}
%     \label{fig:pinned_line}
% \end{figure}

\subsection{Optimal DR-Plan for Pinned Line-Incidence Systems}

%Recall the definitions in Section~\ref{priliminary}.

In this section, we will adapt the results in Section~\ref{sec:DRP} to give the canonical DR-plan for pinned line-incidence graphs. First, we note that
%unlike bar-joint graphs,
\vemph{an isostatic pinned line-incidence graph can be disconnected}, being the disjoint union of two or more isostatic subgraphs. This is because the pins have fixed positions on the plane.
%
%the following difference of pinned line-incidence graphs from the bar-joint graphs.
%
%For pinned line-incidence graphs,
%since the pins have fixed positions on the plane,
%{\em a wellconstrained vertex-maximal proper subgraph can be disconnected},
%being the disjoint union of two or more wellconstrained subgraphs.
%
%\begin{itemize}
%
%\item For pinned line-incidence graphs,
%since the pins have fixed positions on the plane,
%no trivial motion exists, therefore {\em no trivial overconstrained graphs exist} (the smallest overconstrained graph has $6$ vertices).\\
%In particular, note that both a single vertex and a single edge are underconstrained graphs.
%
%\item
%For pinned line-incidence graphs,
%since the pins have fixed positions on the plane,
%{\em a wellconstrained vertex-maximal proper subgraph can be disconnected},
%being the disjoint union of two or more wellconstrained subgraphs. \\
%%Without loss of generality,
%%we slightly modify the requirement of DR-plan for pinned line-incidence graphs
%%such that only connected subgraphs can be considered as a node of the DR-plan.
%
%\end{itemize}
%
We define a \dfn{trivial} graph to be a single vertex and make the following modification to the definition of the canonical DR-plan:

%\begin{itemize}
%\item For any $C$, its $N$ children, $C_1, \ldots, C_N$, are {\em edges} or {\em connected} wellconstrained vertex-maximal proper subgraphs of that node $C$.
%\item A node containing a single edge is a leaf.
%%\item
%%\todo{We just define the %trivial graph/
%%DR-plan leaves for 2-dimensional pinned line-incidence graphs to be single vertices...  }
%%\item \todo{A single edge can be a node of the DR-plan}
%%\item only connected subgraphs can be considered as a node of the DR-plan.
%\end{itemize}

\begin{definition}
    The \dfn{DR-plan} of a  pinned line-incidence graph $G$ is one in which (1) each child node of a non-leaf node $C$ is either a \dfn{connected} rigid vertex-induced subgraph of $C$, or an edge not contained in any  proper rigid subgraph of $C$, and (2) a leaf node is a single edge.
    % Definition~\ref{def:drp} with  modified rules number 2 and 4:
    %    The \textbf{decomposition-recombination (DR-) plan} of  a 2-dimensional pinned line-incidence graph $G$, $DRP(G)$, is  a tree that has the following properties:
    %    \begin{enumerate}
    %        %\item The root of the tree `contains' $G$.
    %        \item[2] For any non-leaf node $C$, each of its children, $C_1,\ldots,C_N$, is either an edge not contained in any wellconstrained proper subgraph of $C$, or a {\em connected} wellconstrained vertex-maximal proper subgraph of  $C$.
    %        %\item The vertex set of $\bigcup_{i=1}^N{C_i}$ is the vertex set of $C$.
    %        \item[4] A node containing a single edge is a leaf.
    %    \end{enumerate}

    The \dfn{canonical DR-plan}  of $G$  is one in which the  child rigid subgraphs are connected, isostatic vertex-maximal subgraphs of the parent.
\end{definition}

Theorem~\ref{theorem:main} holds for pinned line-incidence graphs with this modified definition. The proof is similar to the original proof (in Section~\ref{sec:DRP}) using the same set of lemmas and
%
%\begin{corollary}\label{cor:pinned}
%Given a isostatic pinned line-incidence graph, for node $C$ in $OptimalDRP(G)$ and the children of $C$ in $CompleteDRP(C)$ labeled as $C_1,\ldots,C_N$
%\begin{enumerate}
%    \item if $C_i \cap C_j$ is trivial then all $C_1,\ldots,C_N$ are children of $C$ in $OptimalDRP(G)$.
%    \item if $C_i \cap C_j$ is isostatic then any two out of $C_1,\ldots,C_N$ will be the only children of $C$ in $OptimalDRP(G)$.
%\end{enumerate}
%\end{corollary}
%
the following modified version of Observation~\ref{lemma:union_intersection},
% observation, which
%serves a similar function to Observation~\ref{lemma:union_intersection} in Section~\ref{sec:DRP}
%Using similar counting-based argument, %  to Lemma~\ref{t-l1}--\ref{iuc-l1} in Section~\ref{sec:union_intersection},
which  can be proved using a simple counting based argument.

\begin{observation}\label{lem:pinned_union_intersection}
    Let $F_i$ and $F_j$ to be subgraphs of the same isostatic graph $F$, where each of them can be either a single edge or a connected isostatic subgraph. There are only two possible cases:
    (1) at least one of $F_i$, $F_j$ is an edge, if and only if $F_i \cup F_j$ is underconstrained, if and only if $F_i \cap F_j$ is trivial; and (2) both $F_i$ and $F_j$ are isostatic, if and only if $F_i \cup F_j$ is isostatic, if and only if $F_i \cap F_j$ is isostatic.
    %The following table summarizes all bijections:
    %\setlength{\tabcolsep}{3pt}
    %{\small
    %\begin{center}
    %\begin{tabular}{|p{2.8cm}|c|p{2.3cm}|}
    %\hline
    %\textbf{Property of $F_i,F_j$} & \textbf{$F_i\cup F_j$} & \centering \textbf{$F_i\cap F_j$}  \tabularnewline \hline
    % \centering--- &  \sout{overconstrained}          &  \centering --- \tabularnewline \hline
    %at least one of $F_i,F_j$ is an edge & underconstrained       &  \centering a vertex (underconstrained)  \tabularnewline \hline
    %both $F_i$ and $F_j$ are wellconstrained & wellconstrained        &  \centering wellconstrained  \tabularnewline \hline
    % \centering --- & ---                     &  \centering \sout{underconstrained} \tabularnewline \hline
    %\end{tabular}
    %\end{center}
    %}
    %\setlength{\tabcolsep}{6pt}
\end{observation}

Given Observation~\ref{lem:pinned_union_intersection}, % and the modified definition of DR-plan,
Lemma \ref{lemma:combined_lemma}, Point \ref{lemma:wc_intersection_is_C} and \ref{lemma:uc_intersection_makes_all_uc}, straightforwardly extend to pinned line-incidence graphs. The proof of Point~\ref{lemma:wc_intersection_makes_all_wc} for pinned line-incidence graphs is given in~\ref{sec:appendix_pinned}.
%The following lemma serves as Lemma~ for pinned line-incidence and can be proved similarly.
%\begin{lemma}
%\label{lem:vertex_intersection}
%If $C_i\cap C_j$ is a single vertex, then $\forall k: C_i\cap C_k$ is a single vertex.
%\end{lemma}
%Given Lemma~\ref{lemma:wc_intersection_makes_all_wc} and Lemma~\ref{lemma:uc_intersection_makes_all_uc},
Thus it is straightforward to adapt the proof of Theorem~\ref{theorem:main} to  pinned line-incidence graphs.
%thus proving Corollary~\ref{cor:pinned}.
%
Consequently, we can efficiently find the optimal DR-plan for  pinned line-incidence graphs using basically the same algorithm as for  bar-joint graphs.

% \noindent
% \note The recombination problem for  pinned line-incidence systems is trivial. Since the pins are given fixed positions in the plane, the solutions of a isostatic sub-system will automatically be consistent with solutions of the remaining of the system.

\sidenote{The recombination problem for  pinned line-incidence systems is trivial. Since the pins are given fixed positions in the plane, the solutions of a isostatic sub-system will automatically be consistent with solutions of the remaining of the system.}

\section{Conclusion}
\label{sec:conclusion}

We have clarified the main source of complexity for the optimal DR-plan and recombination problems. For the former problem, when there are no overconstraints (as is the case for 2D qusecs whose realizations are many common types of layered materials), we defined a canonical DR-plan and showed that any canonical DR-plan is guaranteed to be optimal, a strong Church-Rosser property. This gives an efficient (\candrpcomplexity) algorithm to find an optimal DR-plan that satisfies other desirable characteristics.

We have also described a novel method of efficiently realizing a 2D qusecs from the optimal DR-plan by modifying the otherwise indecomposable systems  at nodes of a DR-plan. These results rely on a recent theory of convex Cayley configuration spaces. Relationships and reductions between these and previously studied problems were formally clarified.

We then modeled specific layered materials using extensions of the above theoretical results including the motivating Examples 1-5 in the introduction.

% \section{Open Problems}
\subsection{Open Problems}
\label{sec:appendix:b}
\label{sec:futurework}
\label{sec:open}

The first set of problems are from Section \ref{sec:DRP}:
\begin{openproblem}
    Is there a more efficient algorithm than \candrpcomplexityv\ to find the canonical DR-plan of isostatic 2D bar-joint graphs?
\end{openproblem}

\begin{conjecture}
\label{conj:mfaisoptimal}
    The Modified \frontier\ Algorithm (MFA)~\cite{lomonosov2004graph} finds a canonical, and hence optimal, DR-plan.
\end{conjecture}

The difficulty of proving Conjecture~\ref{conj:mfaisoptimal} arises from the fact that MFA, although running in time $O(n^3)$, is a bottom-up algorithm, involving complex datastructures. However, a proof of optimality, even if it exists, would not be possible without the new notion of a canonical DR-plan at hand. The intuition for this conjecture comes from the similarity of the DR-plan generated by MFA to that of the sequential decomposition described in the proof of Theorem~\ref{theorem:algo_complexity}.  Since it is known \cite{lomonosov2004graph} that the DR-plan generated by MFA is cluster-minimal, an alternate conjecture is the following.

\begin{conjecture}
\label{conj:mfaisoptimal:rephrase}
For independent graphs, cluster-minimal DR-plans are optimal. In fact, for independent graphs, cluster-minimality and canonical are equivalent properties of a DR-plan.
\end{conjecture}

\begin{openproblem}
\label{open:sparsitymatroid}
Although generic rigidity is a property of graphs, and moreover,  in the case of qusecs, generic rigidity has a combinatorial sparsity and tightness-based characterization, the original definition of independence in the rigidity matroid requires an algebraic notion of independence of vectors of indeterminates over $\RR$. Thus the definition of the DR-plan requires algebra over the reals.
In fact, the recursive decomposition problem is not tied to geometric constraint graphs or an algebraic-geometric or mechanical notion of rigidity, and can be defined for any graph using the notion of an abstract rigidity matroid~\cite{graver93book}. This is a type of matroid with two additional matroid axioms; abstract rigidity matroids can be defined in a purely graph-theoretic manner,  with no need for algebra in their definition. However, such abstract rigidity need not have a sparsity characterization. On the other hand, there are sparsity matroids that do not correspond to any notion of abstract rigidity.
However, when an abstract rigidity matroid is also a sparsity matroid, then the techniques of this paper directly apply and we can obtain purely combinatorially defined recursive decompositions of graphs.
\end{openproblem}

A few natural  open questions concern the following common theme that runs through the optimal recombination and later sections of the paper:

\begin{openproblem}
    For fixed $k$, we have
    polynomial time optimal DR-planning (Section~\ref{sec:DRP}),
    recombination (modification) in the presence of $k$ overconstraints,
    optimal modification for decomposition OMD$_k(G)$ when at most $k$ constraints are removed (Section~\ref{sec:recomb}),
    and also optimal completion using at most $k\le 2$ constraints in the body-pin and triangle-multipin cases for a somewhat different optimization of the DR-plan (Section~\ref{sec:table}).
    However, in the running time of all of these algorithms, $k$ appears in the exponent. Can $k$ be removed from the exponent?
\end{openproblem}

One problem in the above theme is from Section \ref{sec:bodypin}.
\begin{openproblem}
    What is the complexity of the optimal completion problem when the given graph has more than 2-dofs? Our proof for the 1 and 2-dof cases relied heavily on the matroidal properties of their corresponding $(k,l)$-tightness. For higher number of dofs, the $(k,l)$ characterization is no longer matroidal \cite{Lee:2007:PGA}. As a result, the major obstacle is that there is no easy way of obtaining an optimal or canonical $k$-dof DR-plan in general. Even assuming such  a DR-plan is available, if higher dofs had the same characteristics, Observation \ref{obs:algebraic_completion} raises questions about the correct measure of DR-plan size that captures algebraic complexity for recombining graphs with many dofs (this is not an issue in the isostatic case). Unless some restrictions can be found and taken advantage of, the $k$-dof optimal completion problem would  have complexity exponential in $k$.
\end{openproblem}

Another problem from the above theme is from Section \ref{sec:recomb}
\begin{openproblem}
    What is the complexity of the restricted OMD (optimal modification for decomposition) problem? This has the potential to be difficult. For example, when the isostatic completion is required to be a 2-tree the restricted OMD problem is reducible to the maximum spanning series-parallel subgraph problem shown by \cite{cai1993spanning} to be NP-complete even if the input graph is planar of maximum degree at most 6. However, since the OMD problem has other input restrictions such as not having any proper isostatic subgraphs, it is not clear if the reverse reduction exists and hence it is unclear whether the OMD problem is NP-complete.

    The same holds for the restricted OMD problem where the isostatic completion is required to be a tree-decomposable graph of low Cayley complexity (i.e.\ have special, small DR-plans). One potential obstacle to an indecomposable graph $G$'s membership in the restricted OMD$_k$ for small $k$ is if $G$ is tri-connected and has large girth. In fact, 6-connected (hence rigid) graphs with arbitrarily large girth have been constructed in \cite{servatius2000rigidity}.
\end{openproblem}

The next is the reverse direction of Observation \ref{obs:OC_to_OMD} in Section \ref{sec:table}.
\begin{openproblem}
    Is the OMD (optimal modification for decomposition) problem reducible to the OC (optimal completion) problem?
\end{openproblem}

More general problem directions are the following.
\begin{openproblem}
    Combinatorial rigidity for periodic structures is an active area of research. This paper motivates a study of the rigidity of self-similar structures, with self-similar groups replacing periodic groups.
\end{openproblem}

\begin{openproblem}
    The pinned line incidence structures of Section \ref{sec:pinnedline}, for example in the case of collagen microfibrils, whose function is elastic contraction, should be considered congruent under projective transformations. I.e.\ the projective group should be factored out as a trivial motion in a new project for extending the combinatorial rigidity characterization of such systems. (Currently we permit no trivial motions at all).
\end{openproblem}

\begin{openproblem}
    Experimental validation (either computationally or physical experiments) of predications based on the material model and theory used in this paper. This can be done by modeling known materials and putting stresses on them, seeing if the prediction is observed in the real material. Or, our theory could be used to design new materials, which can be tested to see if they possess predicted properties.
\end{openproblem}

\appendix
\section{Proofs}
\label{sec:appendix:a}
\label{sec:appendix:proofs}

\subsection{Proofs from Section \ref{sec:DRP}}

% \subsubsection{Proof of Theorem \ref{theorem:main}}
% \label{theorem:main:proof}

\subsubsection{Proof of Observation \ref{lemma:union_intersection}}

\begin{proof}
For (1), simply note that if $F_i\cup F_j$ were trivial, then, by definition, $F_i$ and $F_j$ must be trivial.

For the next parts, we use the quantity $d(G)=2|V|-|E|$, which we call \dfn{density}.
For (2), observe that underconstrained subgraphs of isostatic graphs must have density less than 3.
For (3), observe that, given an isostatic graph, a subgraph with density $3$ must also be isostatic. Then, use the fact that, by definition, $d(F_i)=3$ and $d(F_j)=3$. Then it is straightforward application of the inclusion-exclusion $d(F_i\cap F_j)=d(F_i)+d(F_j)-d(F_i\cup F_j)$.

For (4), because subgraphs of a isostatic graph can only be trivial, underconstrained, or isostatic, all cases have already been exhausted.
\end{proof}

% \begin{proof}
% Use the fact that, by definition, $d(F_i)=k$ and $d(F_j)=k$. Also, the equation $d(F_i\cap F_j)=d(F_i)+d(F_j)-d(F_i\cup F_j)$.
% \begin{enumerate}
%     \item If $F_i\cup F_j$ were trivial, then, by definition, $F_i$ and $F_j$ must be trivial.

%     \item Observe that  Then it is straightforward application of the equation above.

%     % \item Observe that, given a isostatic graph, an underconstrained subgraph must have density less than $k$. Then it is straightforward application of the equation above.

%     \item Observe that, given a isostatic graph, a subgraph with density $k$ must also be isostatic. Then it is straightforward application of the equation above.

%     % \item \textit{Forward direction:} Since $F_i\cup F_j$ is underconstrained and a subgraph of isostatic $F$, it must be that $d(F_i\cup F_j)=l<k$. Therefore $d(F_i\cap F_j)=2k-l>k$. This means $F_i\cap F_j$ is trivial. \textit{Reverse direction:} We know that $d(F_i\cap F_j)>k$ because it is trivial. By the same math, we find that $d(F_i\cup F_j)<k$, showing it is underconstrained.

%     % \item \textit{Forward direction:} We have that $d(F_i\cup F_j)=k$, therefore $d(F_i\cap F_j)=k$. Being a subgraph of isostatic $F$, $F_i\cap F_j$ is also isostatic. \textit{Reverse direction:} By the same math, we find that $d(F_i\cup F_j)=k$ and, since it is also a subgraph of $F$, it is isostatic.

%     \item Subgraphs of a isostatic graph can only be trivial, under-, or isostatic. All cases are already exhausted.
% \end{enumerate}
% \end{proof}

\subsubsection{Proof of Lemma \ref{lemma:combined_lemma}, Point \ref{lemma:wc_intersection_is_C}}

\begin{proof}
Assume $C_i\cup C_j \neq C$. This would contradict the proper vertex-maximality of $C_i,C_j$.
In the reverse direction, we know $C$ is either a non-leaf node (isostatic by definition of a DR-plan) or $G$ itself (isostatic by definition of the problem). Thus, $C_i\cup C_j=C$ is isostatic.
\end{proof}

\subsubsection{Proof of Lemma \ref{lemma:combined_lemma}, Point \ref{lemma:wc_intersection_makes_all_wc}}

\begin{observation}\label{observation:no_edges_between_diff}
Take $R_i=C\setminus C_i$ and $R_j=C\setminus C_j$. If $C_i\cup C_j$ is isostatic, then there can be no edges in $C$ between the vertices of $R_i$ and $R_j$.
\end{observation}

\begin{proof}
Lemma \ref{lemma:combined_lemma}, Point \ref{lemma:wc_intersection_is_C}, shows that $C_i\cup C_j$ must equal the parent graph $C$.
\end{proof}

\begin{proof} (Of Lemma \ref{lemma:combined_lemma}, Point \ref{lemma:wc_intersection_makes_all_wc})

The alternative phrasing is as in Lemma \ref{lemma:combined_lemma}, Point \ref{lemma:wc_intersection_is_C}.

Take $R_i=C\setminus C_i$, $R_j=C\setminus C_j$, and $D_{i,j}=C_i\cap C_j=(C\setminus R_i)\setminus R_j$ (note that $R_j\subset C_i$, $R_i\subset C_j$ and $D_{i,j}\subset C_i,C_j$). Furthermore, take the proper subgraphs $R'_i\subset R_i$, $R'_j\subset R_j$, and $D'_{i,j}\subset D_{i,j}$ that are non-empty.

Assume that there is a third isostatic vertex-maximal proper subgraph $C_k$ (with $C'_k=C\setminus C_k$). There are $3\times 3\times 3 = 27$ possible cases for what this subgraph could be.

Without loss of generality, all graphs are the induced subgraphs of $C$.

% The notation $Idc(G,X)$ is the subgraph of $G$ induced by the vertex set $X\subseteq V$. For a subgraph $H=(W,F)$ of $G$, $Idc(G,H)=Idc(G,W)$.

\newcommand{\inducedOnC}[1]{#1}

\begin{itemize}
    \item 3 cases: $C_k$ cannot be $C=\inducedOnC{R_i\cup R_j\cup D_{i,j}}$, $C_i=\inducedOnC{R_j\cup D_{i,j}}$, or $C_j=\inducedOnC{R_i\cup D_{i,j}}$. This is by definition.

    \item 13 cases: $C_k$ cannot be a proper subgraph of $C_i$ and $C_j$ or else $C_k$ would not be vertex-maximal. These are the graphs $\inducedOnC{R'_i\cup D_{i,j}}$, $\inducedOnC{R'_j\cup D_{i,j}}$, $\inducedOnC{ D_{i,j}}$, $\inducedOnC{R_i\cup D'_{i,j}}$, $\inducedOnC{R_j\cup D'_{i,j}}$, $\inducedOnC{R'_i\cup D'_{i,j}}$, $\inducedOnC{R'_j\cup D'_{i,j}}$, $\inducedOnC{ D'_{i,j}}$, $\inducedOnC{R_i}$, $\inducedOnC{R_j}$, $\inducedOnC{R'_i}$, $\inducedOnC{R'_j}$, and $\inducedOnC{\emptyset}$.

    \item 2 cases: $C_k$ cannot contain $C_i$ or $C_j$ as proper subgraphs, or else they are not vertex-maximal. These are the graphs $\inducedOnC{R'_i\cup R_j\cup D_{i,j}}$ and $\inducedOnC{R_i\cup R'_j\cup D_{i,j}}$ respectively.

    \item 4 cases: \usestwod $C_k$ cannot be $\inducedOnC{R_i\cup R_j}$, $\inducedOnC{R'_i\cup R_j}$, $\inducedOnC{R_i\cup R'_j}$, or $\inducedOnC{R'_i\cup R'_j}$ because these are all disconnected (Observation \ref{observation:no_edges_between_diff}) and cannot be isostatic.

    \item 1 case: $C_k=\inducedOnC{R'_i\cup R'_j\cup D_{i,j}}$ is not possible. Since $C_i\cup C_k = \inducedOnC{R'_i\cup R_j\cup D_{i,j}}\neq C$ we have from Lemma \ref{lemma:combined_lemma}, point \ref{lemma:wc_intersection_is_C}, that $C_i\cup C_k$ cannot be isostatic. We also know it cannot be trivial because it contains isostatic subgraphs. This means it must be underconstrained. From Observation \ref{lemma:union_intersection}, we know that $C_i\cap C_k=\inducedOnC{R'_j\cup D_{i,j}}$ must then be trivial. This is impossible because $D_{i,j}$ is isostatic, thereby contradicting the assumption that $C_k$ is isostatic.

    \item 1 case: \usestwod $C_k=\inducedOnC{R'_i\cup R'_j\cup D'_{i,j}}$ is not possible. Since $C_i\cup C_k\neq C$ (and $C_j\cup C_k\neq C$), we know by the same logic as the previous case that the $C_i\cap C_k$ must be trivial (a single node). However, $C_i\cap C_k=\inducedOnC{R'_j\cup D'_{i,j}}$. This causes a contradiction, the intersection cannot be trivial because $R'_j$ and $D'_{i,j}$ are not empty sets and are disjoint.

    \item 2 cases: \usestwod $C_k=\inducedOnC{R'_i\cup R_j\cup D'_{i,j}}$ and $C_k=\inducedOnC{R_i\cup R'_j\cup D'_{i,j}}$ are not possible. The proof mirrors the previous case, except here you must choose $C_i$ and $C_j$ respectively.

    \item 1 case: $C_k=\inducedOnC{R_i\cup R_j\cup D'_{i,j}}$ is all that remains.
\end{itemize}

Since $D_{i,j}\subset C_i, C_j$ it means that $C_k\cup C_i = C_k \cup C_j = C$, thus proving the Lemma.
%
% \todo{Need a figure?}
\end{proof}

\subsubsection{Proof of Lemma \ref{lemma:combined_lemma}, Point \ref{lemma:uc_intersection_makes_all_uc}}

\begin{proof}
Assume there is some $k$ such that $C_i\cap C_k$ is not trivial. By Observation \ref{lemma:union_intersection}, $C_i\cap C_k$ must be isostatic. Then, by Lemma \ref{lemma:combined_lemma}, point \ref{lemma:wc_intersection_makes_all_wc}, the intersection between any two children must be isostatic. This means that $C_i\cap C_j$ is isostatic. Therefore, such a $k$ cannot exist and all intersections are trivial.
\end{proof}

% \subsubsection{Proof of Theorem \ref{theorem:algo_complexity}}

\subsection{Proofs from Section \ref{sec:recomb}}

% \subsubsection{Proof of Theorem \ref{theorem:omdk}}

\subsubsection{Proof of Theorem \ref{theorem:criterionc}}
\begin{proof}
    The Cartesian realization space of $(H,\left<\delta_{E'}, \lambda_F\right>)$ is computed easily with a DR-plan of size 2, and is the union of $2^t$ solutions (modulo orientation preserving isometries) each with a distinct orientation type, where $t$ is the number of triangles in the 2-tree $H$; here $\delta_{S}$ is the restriction of the length vector $\delta$ to the edges in $S$. A desired solution $p$ (or connected component of a solution space) of $(G,\delta)$ of an orientation type $\sigma_p$ can be found by a subdivided binary search of the Cartesian realization space of $(H, \left<\delta_{E'}, \lambda_F\right>)$ of orientation type $\sigma_p$, as $\lambda_F$ ranges over the discretized convex polytope $\Phi_F(G',\delta_E')$ with bounding hyperplanes described in Theorem \ref{theorem:convexcayley}. A solution $p$ is found  when the lengths for nonedges in $D$ match $\delta_D$.
\end{proof}

\subsection{Proofs from Section \ref{sec:bodypin}}

\subsubsection{Proof of Remark \ref{rem:bodypin_is_barjoint}}
\begin{proof}
    We can replace each body that has only one pin by a single vertex. A body with 2 pins can be replaced by an edge. In general, a body with $n$ pins can be replaced by a 2-tree on $n$ vertices. When finding a DR-plan, we treat each body as trivial, so they become the leaves of the DR-plan. The optimal recombination problem and approach of Section \ref{sec:recomb} are unchanged. The optimal completion via the optimal modification problem in Section \ref{sec:recomb} now has an additional restriction that all edges in the 2-tree representation of the bodies must be removed together, not individually.
\end{proof}

\subsubsection{Proof of Observation \ref{obs:bodypin_drp}}
\begin{proof}
    The existence of this canonical DR-plan follows from the same arguments as in the proof of Theorem \ref{theorem:main}. The only difference is the definition of a trivial intersection. In this case, when two subgraphs share more than 1 body, they become rigid (in fact over constrained). Sharing a pin is not considered an intersection. Such a structure is viewed as two subgraphs each sharing 1 body with a third 1-dof subgraph which essentially just consists of those two bodies pinned together.
\end{proof}

\subsubsection{Proof of Theorem \ref{thm:1dofcase}}
\begin{proof}
    Suppose we are given a body-pin graph and its corresponding body-bar graph $G$ and have obtained the 1-dof DR-plan $T$. Each node of $T$ is then a vertex-maximal proper 1-dof subgraph of $G$.

    To make the graph isostatic, we need only add one body and pin it to 2 other bodies. Doing so will cause $G$ to become $(3,3)$-tight.

We adopt the following algorithm. Choose the 2 bodies to pin to by choosing a node $b$ in $T$ and looking at its children. From Observation \ref{rem:1dofcanon}, we know that the children can only share  a single pin or a sub-graph. Pin the new body to bodies in two separate children. Doing so will ensure that all children of $b$ will have 1-dof and all ancestors of $b$ (including $b$) will now be isostatic.

    % Then, we can form a valid isostatic DR-plan $T_b$ from $T$. In $T_b$, $fanin(b)$is the number of leaves in the subtree rooted at $b$ because no child of $b$ in $T$ is isostatic. Similarly, for any other node $w$ that is an ancestor of $b$, $fanin(w)$ is the the number leaf nodes in the tree rooted at $w$, excluding the subtree constaining $b$, plus 1 (for the node leading to $b$).  Then, for any node $b$ that we choose, $T_b$ is a valid DR-plan. The size of $T_b$ will just be the maximum fanin of all nodes in $T_b$. Thus, if we want to minimize the size of our DR-plan, we simply need to take the $b$ that has the $T_b$ of smallest size.

    Such a pinning covers all possible ways of adding a new body. Assume a new body $b$ is added to the input  graph and pin it to $b_i$ and $b_j$ to make it isostatic. Then, there is a lowest 1-dof node $v$ in $T$ such that $b_i$ and $b_j$ appear in $v$. Thus, pinning $v$ in the manner described yields an equivalent isostatic DR-plan to pinning $b$ to $b_i$ and $b_j$.

    For each node $b$, assign a size of the $T_b$ denoted $|T_b|$. $|T_b| = \displaystyle\max_{v \in T_b} fanin(v)$. We are looking for $b$ that minimizes $|T_b|$. Denote the sub-tree of $T$ rooted at $v$ by $T^v$ and the number of leaves in a tree $T$ by $nl(T)$. Note that $fanin(b)= nl(T^b)$ because no descendant of $b$ is isostatic. Similarly, for any ancestor $w$ of $b$, $fanin(w) = nl(T^w)-nl(T^{b'})+1$, where $b'$ is the child leading to $b$. All other nodes are not isostatic and hence do not appear in the isostatic DR-plan.
    % Then, we can form a valid isostatic DR-plan $T_b$ from $T$. In $T_b$, $b$'s children are now all of the leaf nodes of the subtree rooted at $b$ because no child of $b$ in $T$ is isostatic. Similarly, for any other node $w$ that is an ancestor of $b$, $w$'s children will be the node that leads to $b$, denoted $b'$, along with all of the other leaf nodes in the tree rooted at $w$, excluding $b'$. Then, for any node $b$ that we choose, $T_b$ is a valid DR-plan. The size of $T_b$ will just be the maximum fanin of all nodes in $T_b$. Thus, if we want to minimize the size of our DR-plan, we simply need to take the $b$ that has the $T_b$ of smallest size.

    The node to be pinned is always the the deepest nontrivial node of some path in $T$. Suppose a node $b$ is pinned that has a nontrivial child $v$. Then, $fanin_b(b) = nl(T^b) = nl(T^v) + n$, where $n$ is essentially the number of leaves between $b$ and $v$. If we had instead chosen to pin $v$, then $fanin_v(b) = nl(T^b) - nl(T^{b'}) + 1 \leq fanin_b(v)$. And for each ancestor $w$ of $b$, $fanin(w)$ is unchanged, meaning $|T_v| \leq |T_b|$. Thus we only have to check the deepest non-trivial nodes.

    Running the above algorithm brute force  gives running time quadratic in the number of bodies of the given body-pin system.

    For the multi-triangle pin graphs, we can do the same thing, except we need to add a single triangle to one of the nodes to cause it to become isostatic.
    % Don't leave a blank line between last paragraph and end!
    % For the 2-dof case, we can do something very similar, except instead of pinning a single body 2 times to a node, we can pin another body 2 times to a node. These can be the same node, and if it is the same node, we can find a wellconstrained DR-plan in quadratic time, we would just be doing the same thing as the 1-dof case.
\end{proof}

\subsubsection{Proof of Observation \ref{obs:2dof_case}}
\begin{proof}
    The only difference from the 1-dof case is that now we need to remove
2-dofs from our graph. Start with a 2-dof DR-plan $T$. Like in the previous proof, we need to add a body and 2 pins to 2 nodes to obtain an isostatic DR-plan.

    Suppose we pin 2 distinct nodes $v_i$ and $v_j$. Then, there must exist a common ancestor $a$ of $v_i$ and $v_j$. Then, in $T_{v_i,v_j}$, $fanin_{v_i,v_j}(a) = nl(T^a)$. However, if we chose to pin one of $v_i$ and $v_j$ twice, then $fanin_v(a) = nl(T^a) - nl(T^{a'}) +1$ . Thus $fanin_v(a)' \leq fanin_{v_i,v_j}(a)$. All ancestors of $a$ are unchanged. So $|T_v| \leq |T_{v_i,v_j}|$.

    Thus the only choice is to pin a single node twice. Hence, we can run the same algorithm as the 1-dof case and simply pin twice instead of once.
\end{proof}

\subsubsection{Proof of Observation \ref{obs:algebraic_completion}}
\begin{proof}
    An isostatic graph has 3 parameters that define its position and orientation. These are the Euclidean motions. A 1-dof graph has  4 parameters: the 3 Euclidean motions and a dof parameter. A 2-dof has 5 parameters. The number of parameters roughly correlates with the algebraic complexity of obtaining a realization.

    Thus, starting with a $T$ as described in the proof for Remark \ref{thm:1dofcase}, when a node $b$ is pinned, the same structure is preserved as before. Suppose  $v$ is an isostatic node after pinning $b$. Then, the children of $v$ (except one if $v \neq b$) have 1-dof. The realization complexity for $v$ is simply that of realizing each of its children. In general, the number of parameters for $v$  will be $np(v) = 4nc_1(v)+3$, if $v \neq b$ and $np(b) = 4nc_1(b)$, where $nc_k(v)$ is the number of $k$-dof children of $v$.

    Minimizing the algebraic complexity  requires minimizing the maximum $np(v)$ for any node $v$. In this case, it is not possible to always choose to pin a node closest to a leaf in the tree, because it could have high fan-in. So we  try
brute force by pinning all nodes to  pick the one with the lowest algebraic complexity. This algorithm is still quadratic for the 1-dof case.

    For the 2-dof situation, there are more cases to consider. If we pin the same node twice as above, we have $np(v) = 5nc_2(v)+3$ for any ancestor $v \neq b$ and $np(b) = 5nc_2(b)$. If we pin a node $v$ and one of its ancestors $v'$, then any nodes between $v'$ and $v$ will be 1-dof, any nodes above $v'$ will be isostatic, and nodes below $v$ will be 2-dof. Note that solving or realizing $v'$ will also realize $v$.  Next, we need to consider nodes above and including $v'$ in our complexity: $np(v') = 5nc_2(v') + 4$ and $np(a) = 3 + 5nc_2(a)$ for $a$ an ancestor of $v'$.

The only remaining case is pinning two nodes that are incomparable, i.e.\ do not have a descendant/ancestor relationship. The only change from the previous case is that for the lowest common ancestor of the nodes $v'$, $np(v') = 2*4+5nc_2(v')$. For any ancestor $a$ of $v'$, we still have $np(a) = 3 + 5nc_2(a)$.

    Like the 1-dof case, we again cannot simply choose the nodes deepest in the tree to pin. However, neither can we assume pinning one node twice will give us the best algebraic complexity. Hence, we will need to check each pair of nodes to pin. This makes our brute-force algorithm $O(b^3)$, where $b$ is the number of bodies.
\end{proof}

\subsection{Proofs from Section \ref{sec:pinnedline}}
\subsubsection{Proof of Lemma \ref{lemma:combined_lemma}, Point \ref{lemma:wc_intersection_makes_all_wc} --- for 2-dimensional pinned line incidence graphs}
\label{sec:appendix_pinned}

\begin{proof} %[Proof of Lemma~\ref{lemma:wc_intersection_makes_all_wc} for 2D pinned line incidence graphs]
We use the same notation as in the original proof of  Lemma \ref{lemma:combined_lemma}, Point \ref{lemma:wc_intersection_makes_all_wc}, given above.
Without loss of generality, all graphs are the induced graphs on $C$.

First notice that since $C_i \cup C_j$ is isostatic, by Observation~\ref{lem:pinned_union_intersection}, both $C_i$ and $C_j$ are connected isostatic vertex-maximal proper subgraphs of  $C$. Since $C_j \cup C_j = C$, there are no edges in $C$ that is not contained in a isostatic subgraph, so $C$ does not have any single-edge child node, and $C_k$ is a connected isostatic vertex-maximal proper subgraph of  $C$.

\newcommand{\inducedOnC}[1]{#1}

We analyze all the possible cases for $C_k$.
\begin{itemize}
    \item 1 case: $C_k=\inducedOnC{R'_i\cup R'_j\cup D_{i,j}}$ is not possible. Since $C_k\cup C_i = \inducedOnC{R'_i\cup R_j\cup D_{i,j}}\neq C$ we have from Lemma \ref{lemma:combined_lemma}, Point \ref{lemma:wc_intersection_is_C}, that $C_k\cup C_i$ cannot be isostatic.
    By Lemma~\ref{lem:pinned_union_intersection}, it must be underconstrained,
    so one of $C_i$ and $C_k$ must be an edge, contradicting the assumption that both $C_i$ and $C_k$ are isostatic.

    \item 1 case: $C_k=\inducedOnC{R'_i\cup R'_j\cup D'_{i,j}}$ is not possible. The proof is similar to the previous case.

    \item 2 cases: $C_k=\inducedOnC{R'_i\cup R_j\cup D'_{i,j}}$ and $C_k=\inducedOnC{R_i\cup R'_j\cup D'_{i,j}}$ are not possible.
    The proof is similar to the previous case.
\end{itemize}

All remaining cases are similar to the original proof for 2D bar-joint graphs.
\end{proof}

% \input{appendix_b}

% \begin{figure}\centering
%   \includegraphics[scale=.5]{example.eps}
%   \caption{Example of a normal figure that spans onse column.}
%   \label{fig:square}
% \end{figure}

% \begin{figure*}\centering
%     \includegraphics{example.eps}
%     \caption{Example of a figure that spans both columns.}
% \end{figure*}

%%%%%%%%%%%%%%%%%%%%%%%%%%%%%%%%%%%%%%%%%%%%%%%%%%%%%%%%%%%%%%%%%%%%

\bibliographystyle{elsarticle-num}
\bibliography{paper_references}
% {\scriptsize \bibliography{paper}}

% \clearpage

% \pagenumbering{roman}
% \linespread{1}

\end{document}